\documentclass[aps,pre,preprint,floatfix,amsmath,amssymb]{revtex4-1}
\usepackage{amsmath, amssymb, graphics, mathrsfs, bm, stackrel,bbold}

\usepackage{graphicx}
\usepackage[pdftex,bookmarks={false}, pdfauthor={Srividya Iyer-Biswas}, pdftitle={First Passage  processes in cellular biology}]{hyperref}
\usepackage[all]{hypcap}

\newcommand\ba{\begin{array}}
\newcommand\ea{\end{array}}
\newcommand\nn{\nonumber}

\newcommand\ri{\right}
\renewcommand\le{\left}

\renewcommand\a{\alpha}

\renewcommand\b{\beta}

\renewcommand\d{\delta}

\newcommand\D{\Delta}

\newcommand\G{\Gamma}

\renewcommand\k{\kappa}

\newcommand\m{\mu}

\newcommand\n{\nu}

\newcommand\mbP{\mbs{P}}

\renewcommand\t{\tau}

\renewcommand\th{\theta}

\newcommand\Th{\Theta}

\newcommand\la{\langle}
\newcommand\ra{\rangle}
\newcommand\pd{\partial}
\newcommand\mc{\mathcal}

\newcommand\mbs{\boldsymbol}

\newcommand{\eqwithrate}[1]{\mathrel{\mathop{\rightarrow}\limits^{#1}}}

\begin{document}
\title{First Passage  processes in cellular biology}
\author{Srividya Iyer-Biswas}
\email{iyerbiswas@uchicago.edu}
\affiliation{James Franck Institute and Institute for Biophysical Dynamics,  University of Chicago, Chicago, IL 60637, USA}
\author{Anton Zilman}
\email{zilmana@physics.utoronto.ca}
\affiliation{Department of Physics and Institute for Biomaterials and Biomedical Engineering, University of Toronto, Toronto, ON M5S 1A7, Canada}
\maketitle


\section{Introduction and context.}
Many aspects of the behaviors of physical, chemical and biological systems can be understood simply in terms of the dynamics of the averaged state variables and their deterministic evolution equations. Since such systems typically involve very large numbers of ensemble members, or long time averaging, fluctuations are highly suppressed with respect to the mean behavior. However, for some cases random fluctuations do not simply add negligible noise to the averaged dynamics, instead they give rise to fundamentally different behaviors. The dynamics of the averaged variables are thus insufficient to capture the system's behavior in the stochasticity-dominated regime. Classic examples of systems in this regime include critical phenomena in physics, genetic drift and extinction in biology and diffusion dominated reactions in physical chemistry \cite{feller-book,gardiner-book-2003,Kampen1992,karlin-book-1998,redner-book,Siegert1951,bressloff-book,Montroll1965,weiss1967first,Szabo1980,Kim1958,kimura-1962-fixation}.

It is increasingly appreciated that for biological systems at the cellular and molecular scale, fluctuations can lead to single-cell and single molecule behaviors that considerably deviate from na\"ive ensemble-averaged expectations. This is because the underlying biochemical and biophysical processes often involve reactions with small numbers of reactants. Thus the inherently probabilistic nature of these processes cannot be ignored~\cite{2000-elowitz-lh, 2002-elowitz-sw, 2004-paulsson-fj, 2008-raj-fr, 2010-lestas-lk, 2007-maheshri-zp, Munsky2012}. In addition to this ``intrinsic'' stochasticity, cells may also have additional sources of cell-to-cell variability, known as the ``extrinsic noise''~\cite{2011-hilfinger-pf}. Qualitatively distinct behaviors may emerge when such stochastic fluctuations dominate the system dynamics; familiar biological examples on the cellular scale include stochastic switching between different phenotypes~\cite{2008-acar-zl} and stochastic resonances in neurobiology~\cite{2014-moss-sf}.

Often sharp changes in cellular behavior are triggered by thresholded events, i.e., by the attainment of a threshold value of a relevant cellular or molecular dynamical variable. Since the governing variable itself typically undergoes noisy or stochastic dynamics, there is a  corresponding variability in the times when the same change occurs in each cell of a population. This time is called the ``first passage'' time and the corresponding process is a ``first passage'' (FP) process, referring to the event when a random variable {\em first passes} the threshold value. Even seemingly simple processes, such as the transport of  molecules through channels or multivalent binding also fall under the umbrella of the First Passage processes.

While stochastic effects in copy number fluctuations have received considerable attention in recent years, both experimentally and theoretically \cite{2000-elowitz-lh, 2002-elowitz-sw, 2005-raser-ff, 2004-paulsson-fj, 2005-paulsson-rt,2006-friedman-qd, 2008-raj-fr, 2011-huh-zr, 2010-lestas-lk, 2009-locke, 2007-maheshri-zp, 2009-mukherji-yq, -iyer-biswas-nq, Munsky2012}, the stochasticity in the outcomes and the corresponding noise in the timing of cellular and  molecular events has not received comparable attention. In part, this is due to the experimental challenges in obtaining high quality time series data amenable to analysis for timing noise, which requires making {\em in vivo} measurements at the single cell level~~\cite{-iyer-biswas-kv, 2010-wang-cr}. However, increasingly, this challenge is being overcome through rapid development of single-cell technologies which facilitate making such observations~\cite{2010-wang-cr, -iyer-biswas-kv, 2014-soifer-fh, 2014-kiviet-lj, 2007-talia-tg, 2013-santi-os, 2012-manalis-babymachine, 2013-ouyang-babymachine, 2012-brown-la, 2012-fritzsch-sy, 2011-mir-qf, 2012-moffitt-eg}. On the molecular scale advances in measurement techniques are starting to provide direct insights into the single molecule transport, interactions and signalling processes on the nanoscale \cite{Firnkes2010,wanunu-DNA-review,Tu2013,bezrukov-kullman-maltoporin-2000,winterhalter-2014-chemosensing,grunwald2010vivo,weidong-3d-2013}. These technological developments have made it apposite to now develop the FP formalism specifically for addressing current problems in cellular and molecular biology, i.e., for establishing  quantitative relations between the timing noise in stochastic events and the corresponding underlying stochastic dynamics of the thresholded variables.

Mathematical techniques for modeling and analyzing First Passage processes were pioneered a few decades ago, in the context of non-equilibrium physical chemistry and chemical physics \cite{Siegert1951,Montroll1965,weiss1967first,Szabo1980,Kim1958}. Detailed descriptions can be found in several textbooks and reviews \cite{redner-book,karlin-book-1998,feller-book,gardiner-book-2003,Hanggi2009,jacobs2010stochastic,gillespie-book-1991}. These  techniques are now increasingly being adapted to  problems in cellular, molecular and population biology.  The renewed interest in FP problems in biological contexts has been reflected in several new works summarizing various aspects of the applications of FP theory to these problems \cite{Bressloff-Review-2014,chou-review-2014,bressloff-book,muthukumar-book}. However, many fundamental and practically useful results remain scattered across the literature in somewhat disparate communities. 

In this review we first present and elucidate fundamentals of the FP formalism within a unified conceptual framework, which naturally integrates the existing techniques. We then discuss applications thereof, with emphasis on the practical use of FP techniques in biophysical systems. Our focus here is on covering a diverse set of analytical techniques;  the number of reviewed biological applications is thus limited, out of necessity. We focus on three specific areas: channel transport; receptor binding and adhesion; and single-cell growth and division.
\section{Framework.}
The presentation in this section is partially drawn from these textbooks and reviews: \cite{redner-book,gardiner-book-2003,karlin-book-1998,bressloff-book,chou-review-2014,Bressloff-Review-2014}.
\subsection{Stochastic processes.}
We first review  fundamentals of the theory of stochastic processes. The system dynamics are specified by the set of its states, $\{S\}${,} and the transitions between them, $S\rightarrow S'$, where $S, S' \in \{S \}$. For example, the state $S$ can denote the position of a Brownian particle, the numbers of molecules of different chemical species, or any other variable that characterizes the state of the system of interest. Here we restrict ourselves to processes for which the transition rates depend only on the system's instantaneous state, and not the entirety of its history. Such memoryless processes are known as  Markovian and are applicable to a wide range of systems. We also assume that the transition rates do not explicitly depend on time, a condition known as stationarity. In this review we make the standard assumption that the transitions between the states are Poisson distributed random processes. In other words, the probability of transitioning from state $S'$ to state $S$ in an infinitesimal interval, $dt$, is $\alpha (S,S')dt$, where $\alpha (S,S')$ is the transition rate.

\emph{Examples.} For a  Brownian particle diffusing along a line, the state $S$ is defined by the particle position; the transition rate is $2D/d^2$, where $D$ is the diffusion coefficient and $d$ is the step length. For a set of radioactive atoms undergoing decay with rate $\k$ per atom, the state $S$ is defined by the number, $n$, of atoms that have not decayed yet, and the transition rate from state $n$ to state $n-1$ is $\k \,n$. For a system with $N$ reacting chemical species, the system state is defined by the  concentrations of each reactant, $(x_{1}\ldots x_{N})$, and the transition rates are functions of these concentrations.  


\subsection{Time evolution equation(s) for the system.}
We now summarize the equations that govern the dynamical evolution of the probability that the system is in state $S$ at  time $t$,  which we denote by $P(S,t)$. Typically, such equations are written in one of  three formalisms: the Master Equation, the Fokker-Planck Equation, or the Stochastic Differential Equation; each is summarized below in turn. Details of the derivations can be found in \cite{karlin-book-1998, gardiner-book-2003,Kampen1992}.

\subsubsection{The Master Equation.}\label{sec-forward-equations}
The Master Equation (ME) is the most general of the three formalisms and comprises of a set of linear ordinary differential equations. The ME is derived as follows. The probability of being in state $S$ at a time $t+dt$, $P(S,t+dt)$, is the sum of the following two terms. First, the probability, $P(S,t)$, that the system was already in the state $S$ at time $t$ and remained there during $dt$. Second, the probability that the system was originally in some other state $S'$ at time $t$, times the probability that the system transitioned from $S'$ to $S$ during $dt$. Combining these two terms one obtains
\begin{align}
P(S,t+dt) &= P(S,t)\le[ 1 - \sum_{S'}\a(S',S)dt \ri] + \sum_{S'}P(S',t)\a(S,S')dt\nonumber.
\end{align}
Taking the limit  $dt\to 0$, we get the  \emph{forward} Master Equation (FME), or simply the Master Equation (ME):
\begin{align}\label{eq-masterequation0}
\pd_{t}P(S,t) &= \sum_{S'}\a(S,S')P(S',t) - \sum_{S'}\a(S',S)P(S,t).
\end{align}
The first term in Eq. \ref{eq-masterequation0} is the probability flux into the state $S$ while the second term is the flux out of $S$. The Master Equation can be compactly written in operator notation as
\begin{align}\label{eq-forward-master-operator-form}
\pd_{t}\mbP(t) &= \mc{M}_{f} \mbP(t),
\end{align}
where $\mbP(t)$ is a vector with the components $P(S,t)$ and $\mc{M}_{f}$ is a linear operator with the components
\begin{align}\label{eq-Mforward}
\le(\mc{M}_{f}\ri)_{SS'} &= \a(S,S') - \d_{SS'} \sum_{S''} \a(S'',S).
\end{align}
Note that Eq.~\eqref{eq-masterequation0} conserves probability, $\pd_{t}\le(\sum_{S} P(S,t)\ri) = 0$, which is guaranteed since $\sum_{S}\le(\mc{M}_{f}\ri)_{SS'}=\sum_{S}\a(S,S')- \sum_{S''} \a(S'',S')=0$. For systems with discrete states, the operator  $\mc{M}_{f}$ is simply a matrix with the above components.

Often physical systems  have states that are either characterized by a continuous variable $s$, or can be  conveniently viewed as a continuous limit of the discrete states $S$. In this case,  the  discrete probabilities, $P(S,t)$, are replaced by the probability density, $p(s,t)$, that specifies the probability that the state of the system lies in an infinitesimal region $[s,s+ds]$ in $s$-space: $P([s,s+ds],t) = p(s,t)ds$. The sums in the Master Equation \eqref{eq-masterequation0} are then replaced by the corresponding integrals:
\begin{align}\label{eq-continuous-ME}
\pd_{t}p(s,t) &= \int\a(s,s')p(s',t)ds' - p(s,t) \int\a(s',s)ds'.
\end{align}
Both in the discrete and the continuous cases the formal solution to the Master Equation can be written  in terms of the initial probability distribution, $\mbP(t_{0})$, at the initial time, $t_{0}$, as
\begin{align}\label{eq-linearevolutionoperator}
\mbP(t) &= e^{\mc{M}_{f}(t-t_{0})}\mbP(t_{0}).
\end{align}
For discrete state variables, Eq. ($\ref{eq-linearevolutionoperator}$) simply requires exponentiation of a matrix, whereas for the case of continuous state variables, it generally requires  solution of the integral equation (\ref{eq-continuous-ME}).

\emph{Examples}. In the previously mentioned example of radioactively decaying atoms, the Master Equation is $\pd_{t}P(n,t) = (n+1)\k P(n+1,t) - n\k P(n,t)$, where $\k$ is the decay rate per atom. For a particle performing an unbiased random walk with jump length $a$ and total jump rate $r$, the Master Equation is $\pd_t P(x,t)=\frac{r}{2}P(x+a,t)+\frac{r}{2}P(x-a,t)-rP(x,t)$.


\subsubsection{Kramers-Moyal expansion and the Fokker-Planck equation.}\label{sec-FP-framework}

In some cases where the state space is continuous, there is a sense of locality, and one can define a ``distance" between two states $s$ and $s'$, $\delta=s'-s$.  A familiar example of this scenario is a Brownian particle on a line, its instantaneous state being specified by its coordinate. The  Master Equation, Eq. \eqref{eq-continuous-ME}, then can be rewritten as
\begin{align}
\label{eq-1p6}
\pd_{t}p(s,t)= \int r(\d,s+\d)p(s+\d,t)d\d - p(s,t) \int r(\d,s)d\d,
\end{align}
where $r(\delta,s)\equiv \a(s+\d,s)$ is the rate of jumping over a distance $\delta$, away from the state $s$. If $r(\delta,s)$ rapidly decays with increasing $\d$, over the lengthscale of typical variations of the probability  density, $p(s,t)$, one can expand $r(\d,s+\d)$ and $p(s+\d,t)$ around $\d=0$. Thus $p(s+\delta,t)=p(s,t)+\d\, \pd_{s}{p(s,t)}+ \d^{2} \pd_{s}^{2}p(s,t)/2+ \mc{O}(\d^{3})$ and $r(\delta,s+\delta)= r(\d,s)+\d\, \pd_{s}{r(\d,s)}+\d^{2} \pd_{s}^{2}r(\d,s)/2+ \mc{O}(\d^{3})$; this is known as the Kramers-Moyal expansion. Long-tailed transition rates,  $r(\delta, s)$, result in anomalous diffusion, not addressed in this review \cite{klafter-anomalous-review}. Substituting these expansions in Eq.~\eqref{eq-1p6} and keeping terms  till the second order in $\delta $, the minimum order necessary for obtaining non-trivial diffusion-like motion, we arrive at the following partial differential equation, known in physics literature as the Fokker-Planck Equation:
\begin{align}\label{eq-fokkerplanck0}
\pd_{t}p(s,t) &= - \pd_{s}\le(A(s) p(s,t)\ri) + \frac{1}{2}\pd^{2}_{s}\le(B(s) p(s,t)\ri),
\end{align}
where the functions $A(s)$ and $B(s)$ are, respectively, the first and the second moments of the transition rate $r(\d,s)${:}
\begin{align}
A(s) &= \int_{-\infty}^{\infty} r(\d,s) \d d\d\;,\;\;\; B(s) = \int_{-\infty}^{\infty} r(\d,s) \d^{2} d\d.
\end{align}
As we have noted previously, the total probability is conserved by the Master Equation. Analogousy, the Fokker-Planck equation, Eq.~\eqref{eq-fokkerplanck0}, conserves probability and can be written as a local continuity equation for the probability density,
\begin{align}\label{eq-continuityequation0}
\pd_{t}p(s,t) + \pd_{s}J(s,t) &= 0,
\end{align}
where the quantity $J(s,t) = A(s) p(s,t) - \pd_{s}(B(s)p(s,t))/2$ is the \emph{probability current}. It is important to emphasize that the Fokker-Planck equation is an uncontrolled approximation to the full Master Equation, and can lead to different results \cite{feller-book,Kampen1992}.

\emph{Physical interpretation.} When the Fokker-Planck equation is used to describe the movement of a physical particle under the action of a force $f(s)$, the equilibrium probability density has to satisfy the Boltzmann-Gibbs distribution, $p(s)\propto \exp(-\int^s f(x)dx/k_BT)$. Rewriting the probability current in the Fokker-Planck equation as
\begin{align}
J(s,t)=\tilde{A}(s)p(s,t)-\frac{1}{2}B(s)\pd_{s}p(s,t),
\end{align}
with $\tilde{A}(s)=A(s)-\frac{1}{2}\pd_{s}B(s)$, we see that the equilibrium solution to the FPE, which should satisfy the condition $J(s,t)=0$, is $p(s,t) \propto \exp(\int^s 2\tilde{A}(x)/B(x)dx)$.  Comparing with the Boltzmann-Gibbs distribution, this imposes the constraint  $\tilde{A}(s)=- {B(s)f(s)}/{2k_BT}$, which is known as the Einstein relation. Written this way, the current $J(s,t)$ has a simple physical interpretation: the first term in the current is the drift,  arising due to the action of the force and characterized by a velocity $\tilde{A}(s)$, and the second term represents the diffusive flux (Fick's law) with the diffusion coefficient $D(x)=B(x)/2$.

\subsubsection{Langevin and the Stochastic Differential Equations.}
When the state variable is continuous, the stochastic evolution of the system can be thought of as deterministic motion with added random fluctuations. A familiar example of this case is the Langevin equation that describes the motion of a diffusing Brownian particle,
\begin{align}\label{eq-langevin}
\dot{x}(t)=\mu(f(x)+\xi(t)),
\end{align}
where $f(x)$ is the deterministic force acting on the particle, $\xi$ is a random force that mimics the effects of random jumps, and $\m$ is the mobility. Note that $x(t)$ is now a \emph{random variable}. The formal connection between this representation and the probability density of the previous section is provided by the relation $p(s,t)=\langle \delta (x(t)-s)\rangle_{\xi}$, where the average is over all the realizations of the random force, $\xi$.

It can be shown that with the choice of $\mu=D/kT$ and $\langle \xi(t)\xi(t') {\ra} =\frac{kT}{D^{1/2}}\delta(t-t')$, this equation is mathematically equivalent to the following Fokker-Planck equation,
\begin{align}\label{eq-fokkerplanck-SDE}
\pd_{t} p(s,t) = - \pd_{s}\le( \mu f(s) p(s,t)\ri) + D\pd^{2}_{s} p(s,t).
\end{align}
More generally, any  Fokker-Planck equation of the form
\begin{align}\label{eq-fokkerplanck-B(x)}
\pd_{t}p(x,t) &= - \pd_{x}\le(A(x) p(x,t)\ri) + \frac{1}{2}\pd^{2}_{x}\le(B(x) p(x,t)\ri)
\end{align}
has an equivalent stochastic differential equation (SDE) of the form
\begin{equation}\label{eq-SDE-Ito}
\dot{x}(t)= A_0(x)+B_0(x)\chi(t),
\end{equation}
with delta-correlated random term $\langle\chi(t)\chi(t')\rangle=\delta(t-t')$. However, due to mathematically pathological properties of the function  $\chi(t)$ (it is nowhere differentiable),  when $B_0(x)$  depends on $x$, the Eq. (\ref{eq-SDE-Ito}) is not unambiguosly defined. In general, its interpretation requires re-definition of the rules of differentiation and integration, and many different SDEs can be chosen to correspond to the same FPE, depending on the interpretation.

Historically, the two major interpretations are from Ito and Stratonovich. In both these formulations, $B_0(x)=B(x)^{\frac{1}{2}}$. However, $A_0(x)=A(x)$ in Ito interpretation while $A_0(x)=A(x)-\frac{1}{4}\pd_{x}B(x)$ in the Stratonovich interpretation. From the practical perspective, Ito interpretation allows one to simulate  the SDE using the usual forward Euler scheme. However, special differentiation and integration rules are required for analytical calculations. On the other hand, Stratonovich interpretation allows using the regular rules of calculus but has to be simulated using implicit schemes. We emphasize that the Fokker-Planck equation does not suffer from such ambiguity of interpretation;  SDEs corresponding to different interpretations of the same Fokker-Planck equation lead to the same physical results~\cite{Kampen1992,bressloff-book}.

\subsection{Backward evolution equations.}\label{sec-backward-equations}
\subsubsection{The Backward Master Equation.}
The general form of the Master Equation derived in Eq.~\eqref{eq-masterequation0} is also known as the \emph{forward} Master Equation (FME), since it  describes the evolution from an initial state to a state at a later time.  The Master Equation is linear in the probabilities $P(S,t)$ (see Eq.~\ref{eq-masterequation0}). Thus, its solution with any general initial condition, $P(S,t_{0})$, can be obtained as a linear combination of the conditional probabilities $P(S,t| S_{i},t_{0})$, which are solutions to the Master Equation for the special initial conditions, $P(S,t_{0}) = \d_{S,S_{i}}$. Mathematically, the  $P(S,t| S_{i},t_{0})$ are  the Green's functions of the Master equation.

The time evolution of these conditional probabilities, $P(S,t | S_{i},t_{0})$, can also be described by an alternative linear equation instead of the Forward Master Equation, known as the Backward Master Equation (BME), which is especially useful in the context of First Passage problems. The key to deriving the BME equation is to consider the \emph{first} step out of the initial state $S_i$ at time $t_0$, rather than the \emph{last} step of the trajectory, leading to the  state $S$, at time $t$. Similar to the derivation of the FME, the conditional probability, $P(S,t | S_{i},t_{0})$, can be written down as the sum of the probabilities of two mutually exclusive events: (i) that the system transitioned to a different state $S'$ during  the time interval $ dt$, with the probability $\a(S',S_{i}) dt$, and then evolved to a state $S$ by time $t$, with the probability $P(S,t | S',t_{0}+dt)$, or (ii) that the system was still in state $S_{i}$ at time $t_{0} + dt$, with the probability $1 - \sum_{S'}\a(S',S_{i}) dt$, and then by time $t$  evolved to the state $S$, with the probability $P(S,t | S',t_{0}+dt)$. Together, these terms yield
\begin{align}
P(S,t | S_{i},t_{0}) &= \le(1 - \sum_{S'}\a(S',S_{i}) dt\ri)P(S,t |S_{i},t_{0}+dt) \nn \\
&+ \sum_{S'} \a(S',S_{i}) dt \times P(S,t|S',t_{0}+dt).
\end{align}
The stationarity condition, i.e. the the lack of explicit dependence of the transition rates on time, implies that the conditional probability $P(S,t | S_{i},t_{0})$ is a function only of $t-t_{0}$. Thus, $P(S,t | S',t_{0} + dt) = P(S,t - dt | S',t_{0})$ in the above equation and so
\begin{align}\label{eq-backwardmasterequation0}
P(S,t | S_{i},t_{0}) &= \le(1 - \sum_{S'}\a(S',S_{i}) dt\ri)P(S,t - dt | S_{i},t_{0}) \nn \\
&+ \sum_{S'} \a(S',S_{i}) dt \times P(S,t - dt|S',t_{0}).
\end{align}
Taking the limit $dt\to 0$, we get
\begin{align}
\pd_{t} P(S,t | S_{i},t_{0}) &= \sum_{S'} \a(S',S_{i}) P(S,t | S',t_{0}) - P(S,t | S_{i},t_{0}) \sum_{S'}\a(S',S_{i}) \nn \\
&\equiv \sum_{S'} \le(\mc{M}_{b}\ri)_{S_{i} S'} P(S,t | S',t_{0}).
\end{align}

The resulting equation is known as the Backward Master Equation (BME); $\mc{M}_b$ is the Backward Master Operator with the components $\left(\mc{M}_b\right)_{SS'}=\a(S',S)-\d_{S,S'}\sum_{S'}\a(S',S)$, and it is a transpose of the  the forward operator $\mc{M}_f$ defined in Eq.~\eqref{eq-Mforward}. The Backward Master Equation can be also written in the operator form:
\begin{align}\label{eq-backward-master-operator-form}
\pd_{t}\boldsymbol{P}^T(t) &=\mc{M}_{b}\cdot\boldsymbol{P}^T,
\end{align}
where $\boldsymbol{P}^T$ is a vector whose $i$-th component is $P(S,t|S_i)$: \[\boldsymbol{P}^T=(...P(S,t|S_1),P(S,t|S_2)...P(S,t|S_i)...).\]

\subsubsection{The Backward Fokker-Planck equation.}
The Backward Master Equation can be extended to the continuous case, similar to the procedure applied to the FME in Section \ref{sec-FP-framework}, and can be approximated by the corresponding Backward Fokker-Planck equation,
\begin{align}
\pd_{t}P(x,t | x_{i}, t_{0}) &= A(x_{i})\pd_{x_{i}}P(x,t | x_{i}, t_{0})+ \frac{1}{2}B(x_{i})\pd^{2}_{x_{i}}P(x,t | x_{i}, t_{0}).
\end{align}
It is analogous to the forward Fokker-Planck equation, Eq.~\eqref{eq-fokkerplanck0}, except that the differential operators on the right hand side act on the initial state, $x_{i}$, instead of the current state, $x$, resulting in $B(x_i)$ being outside of the derivative sign. Note that the Backward Fokker-Planck equation conserves the overall probability; however, it cannot be written as a local continuity equation with respect to the initial position $x_i$.

\subsection{First Passage Processes.}\label{sec-FPT-framework}
We have now set up the framework required to address the First Passage (FP)  problem, which can  be stated as the following question. For a stochastic Markov process that starts from the initial state $S_{i}$ at time $t_{0}$, what is the distribution of times, $\tau=t-t_{0} $, at which the system arrives at the specific state $S_{f}$ \emph{for the first time}? We denote the probability density of this First Passage Time distribution by $F(\tau; S_{f} | S_{i})$; due to stationarity it does not depend explicitly on $t_0$.

\begin{figure}[h]
\begin{center}
\resizebox{10cm}{!}{\includegraphics{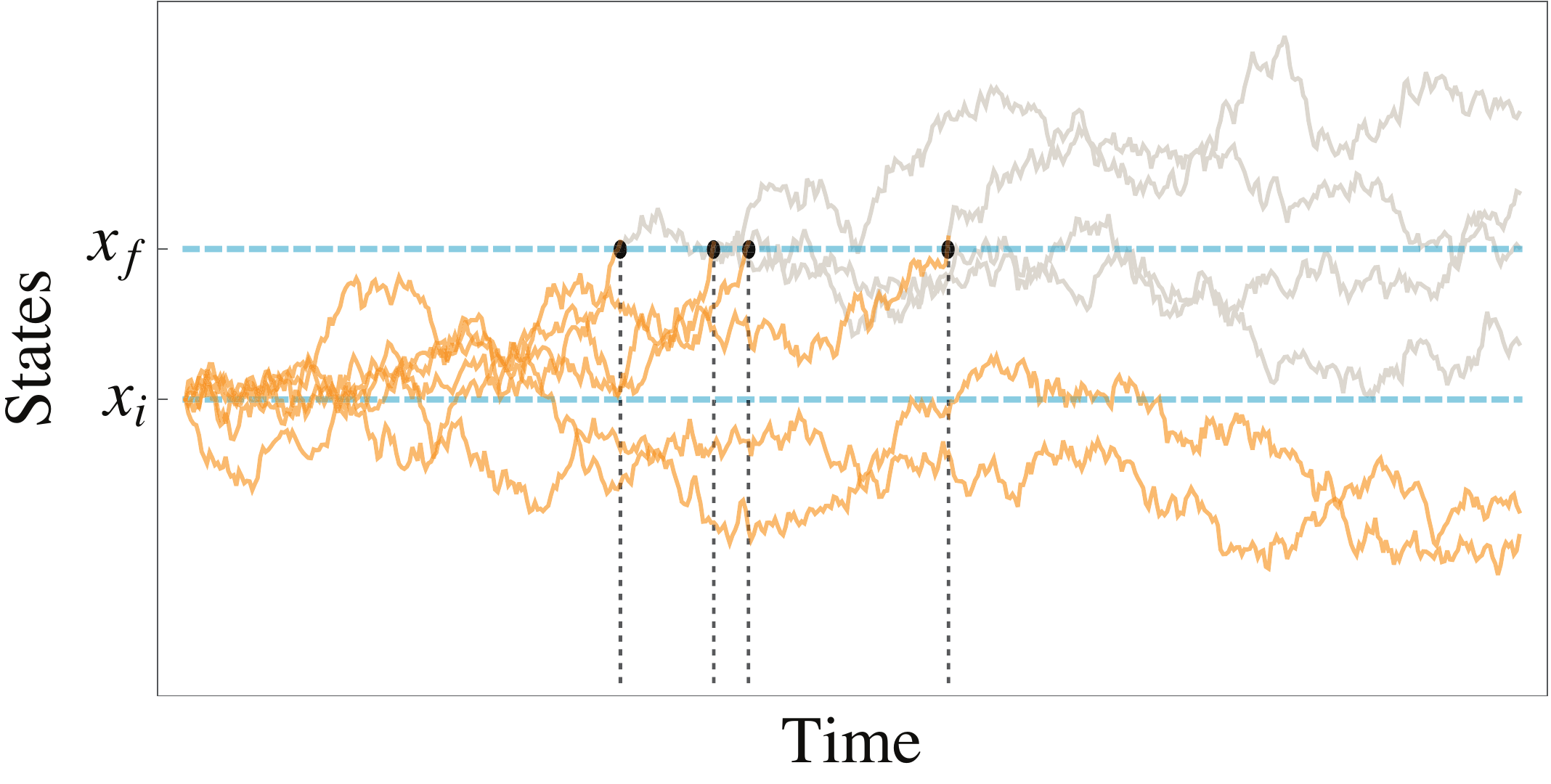}}
\caption{\textbf{The First Passage Time problem.} Distributions of times when a system, starting from an initial state, $x_{i}$, first visits specified threshold, $x_{f}$, can be found by considering an auxiliary problem, with an absorbing boundary condition at $x_{f}$. (See Section \ref{sec-FPT-framework}.) Shown here are 6 sample trajectories starting from $x_{i}$ (orange); the first passage time for each trajectory is marked by the dotted vertical line at the intersection of the trajectory with the threshold, $x_{f}$. In the auxiliary problem, the trajectories continuing beyond the first visitation event (shown in gray) are irrelevant and should not be counted.}
\label{fig-fpt}
\end{center}
\end{figure}

Na\"ively, one may be tempted to guess that the first passage time distribution should be proportional to the probability to be in state $S_f$ at time $t$, $P(S_{f}, t | S_{i}, t_{0})$. However, this is incorrect because $P(S_{f}, t | S_{i}, t_{0})$  contains contributions from trajectories in which the system has already visited the final state $S_{f}$ at other instances between times $t$ and $t_{0}$. In other words, $P(S_{f}, t | S_{i}, t_{0})$ over-counts the number of first passage trajectories. See Fig~\ref{fig-fpt} for a graphic representation of the first passage time problem.

\subsubsection{First Passage Processes using Forward Master and Fokker-Planck equations.}
Counting trajectories that have not visited the final state previously is a combinatorially complex problem. It can be solved by considering an auxiliary version of the original problem, in which once the system arrives at the state $S_{f}$, it remains there indefinitely. Thus it is not allowed subsequent transitions to  another state.
In other words, one places an \emph{absorbing} boundary condition at the state $S_{f}$, with transition rates out of $S_f$, $\a(S,S_{f})$,  being set equal to zero for all $S$.

In this auxiliary problem we define the survival probability $\mc{S}(t, S_{f} | S_{i}, t_{0})$ as the probability that the system has not yet been absorbed at $S_{f}$ by time $t$, after starting from $S_{i}$ at $t=t_{0}$:
\begin{align}
\mc{S}(t, S_{f} | S_{i}, t_{0}) &= \sum_{S\neq S_{f}} P(S,t  | S_{i}, t_{0}).
\end{align}
Note that  stationarity assumption dictates that $\mc{S}$ is a function of $t-t_{0}$ only. Since, by definition, the probability of reaching $S_f$ in a time interval $[t_0+\tau,t_0+\tau+d\tau]$ is $F(\tau,S_f|S_i)d\tau$, the probability of reaching $S_f$ \emph{by} time $t$ is $\int_{0}^{t-t_0} F(\tau;S_f|S_i)d\tau$. In other words, the probability that the First Passage Time is larger than $\tau$ is $\mc{S}(t_0+\tau, S_{f} | S_{i}, t_{0})$, and therefore $\mc{S}(t_0+\tau, S_{f} | S_{i}, t_{0})$  is the cumulative distribution of $F(\tau,S_f|S_i)$. Intuitively, it is clear that the survival probability $\mc{S}(t)$ decreases in time with the rate equal to the probability current into the absorbing state $S_f$:\newline 
$\partial_t\mc{S}(t, S_{f} | S_{i})= -J(S_f,t|S_i)$.  This provides a prescription for obtaining the First Passage Time distribution, $F(S_f,\tau|S_i)$, by solving the Forward Master Equation, which yields the probabilities $P(S,t|S_i)$, and hence the probability flux into the absorbing state.

\emph{Formal derivation}. These arguments can be put in a mathematically rigorous form. The survival probability is related to the FPT distribution as
\begin{align}\label{eq-22}
\mc{S}(t, S_{f} | S_{i}, t_{0}) &= 1 - \int_{0}^{t-t_0} F(\tau;S_f|S_i)d\tau.
\end{align}
Thus, the survival probability is the cumulative probability distribution for $F(\tau;S_f|S_i)$ and
\begin{align}\label{eq-F-S-derivative}
F(\tau; S_{f} | S_{i}) &= -\pd_{t}S(t, S_{f} | S_{i}, t_{0})|_{t=t_0+\tau}.
\end{align}
Using this with the Forward Master Equation, and keeping in mind that $(\mc{M}_f)_{S,S_{f}}=0$  because $\alpha(S,S_f)=0$ for all $S$ (see Eq. \ref{eq-linearevolutionoperator}),
\begin{align}\label{eq-S(t)-from-ME}
\partial_t\mc{S}(t, S_{f} | S_{i}, t_{0}) &= \sum_{S\neq S_f}\partial_t P(S,t|S_i,t_0)=\sum_{S\neq S_f}\sum_{S'\neq S_f}(\mc{M}_f)_{S,S'}P(S',t|S_i,t_0) \nn \\
&= \sum_{S'\neq S_f}P(S',t|S_i,t_0)\sum_{S\neq S_f}(\mc{M}_f)_{S,S'}\nn\\
&=\sum_{S'\neq S_f}P(S',t|S_i,t_0)\le(\sum_{S}(\mc{M}_f)_{S,S'}-(\mc{M}_f)_{S_f,S'}\ri)\nn\\
&=-\sum_{S'}\alpha(S_f,S')P(S',t|S_i,t_0)\equiv -J(S_f,t|S_i,t_0).
\end{align}
We have used the facts that $\sum_{S}(\mc{M}_f)_{S,S'}=0$ due to the conservation of probability and that $\left(\mc{M}_f\right)_{S_fS'}=\alpha(S_f,S')$ (see section \ref{sec-forward-equations}). The quantity $J$ in the last line is the probability current from all accessible states into $S_f$. Comparing  with  Eq. \ref{eq-F-S-derivative}, this proves  our heuristic assertion that $F(\tau; S_{f} | S_{i}) = J(S_f,t_{0}+\t|S_i,t_0)$.

This result can also be obtained for a continuous variable using the Fokker-Planck equation (Eq.~\eqref{eq-fokkerplanck0}), shown below for  a simple one-dimensional case.
Putting $t_0=0$ and assuming $x_i<x_f$ (and thus $p(x>x_{f},t) = 0$),
\begin{align}
&F(t; s_f|s_i) = -\pd_{t}\mc{S}(t,s_f| s_{i})= -\pd_{t}\le(\int_{-\infty}^{s_{f}} p(s,t|s_i)ds \ri)_{t=\tau} \nn\\
&= - \left(\int_{-\infty}^{s_{f}}\pd_{t}p(s,t|s_i)ds\right) = \int_{-\infty}^{s_{f}} \pd_{s}J(s,t|s_i)ds = J(s_f,t|s_i),
\end{align}
since the current at infinity vanishes, i.e., $J(\infty,t|s_i)=0$. 

To summarize, in order to calculate the probability density of the First Passage Times to state $S_f$, from state $S_i$,  one needs to solve the Forward Master or Fokker-Planck equation for the auxiliary process with the absorbing boundary condition at $S_{f}$, obtain the probability current $J(t)$ into the absorbing state $S_f$, which then provides the FPT distribution through the relation $F(\tau; S_{f} | S_{i}) = J(S_f,t_{0}+\t|S_i,t_0)$.


\subsubsection{First Passage Processes using Backward Master and Fokker-Planck equations}
The First Passage Time distribution can  also be calculated using the backward formalism of Eq. \ref{eq-backward-master-operator-form}. The crucial insight is that the survival probability, $\mc{S}$, also satisfies the Backward Master Equation. Setting $t_0=0$,
\begin{align}
&\pd_{t}\mc{S}(t, S_{f} | S_{i}) = - \sum_{S\neq S_{f}} \pd_{t} P(S ,t| S_{i})
= - \sum_{S\neq S_{f}}\sum_{S'} \le(\mc{M}_{b}\ri)_{S_{i} S'} P(S  ,t | S')\nn\\ &=- \sum_{S'}\le(\mc{M}_{b}\ri)_{S_{i} S'}\sum_{S\neq S_f} P( S,t | S_{i}) = - \sum_{S'} \le(\mc{M}_{b}\ri)_{S_{i} S'} \mc{S}(t, S_{f} | S'),
\end{align}
where we have used the linearity of $\mc{M}_{b}$. Taking another time derivative, we find that the FPT probability density also obeys the Backward Master Equation:
\begin{align}
\pd_{t} F(t; S_{f} | S_{i}) &=-\pd_{t}^2 S(t; S_{f} | S_{i})=\sum_{S'} \le(\mc{M}_{b}\ri)_{S_{i} S'} F(t; S_{f} | S').
\end{align}
Thus the First Passage Time distribution can be obtained by solving the Backward Master Equation (and correspondingly, the Backward Fokker-Planck equation).

Although solving the Backward Master Equation is not necessarily easier than solving the Forward Master Equation, it provides a relatively easy way of calculating the moments of the $F(t;S_f|S_i)$. For instance, the Mean First Passage Time (MFPT), defined as
\begin{equation}
T(S_f|S_i)\equiv \int_0^{\infty} d\tau\; \tau F(\tau ; S_{f} | S_{i}),
\end{equation}
can be calculated by applying the Backward Master operator to both sides:
\begin{align}\label{eq-MFPT-backward}
\sum_{S'}(\mc{M}_b)_{S_i S'}T(S_f|S') &=\int_0^{\infty} d\tau\; \t \sum_{S'}(\mc{M}_b)_{S_i S'} F(\t; S_{f} | S') \nn \\
&=\int_0^{\infty} d\t\; \t \pd_t F(\t; S_{f} | S_{i}) =\int_0^{\infty} d\t \pd_\t \mc{S}(\tau, S_{f} | S_{i}, 0) \nn\\
&=\mc{S}(\infty, S_{f} | S_{i})-\mc{S}(0, S_{f} | S_{i})=-1,
\end{align}
where we have used the fact that $\mc{S}(0, S_{f} | S_{i})=1$ and $\mc{S}(\infty, S_{f} | S_{i})=0$. In  matrix form,
\begin{align}
\mc{M}_b\cdot\boldsymbol{T}=-1,
\end{align}
where $\boldsymbol{T}$ is the vector whose $i$-th component is $T(S_f|S_i)$.

Higher moments can be obtained by sequential application of the reasoning of Eq. (\ref{eq-MFPT-backward}).
This obviates the need for solving the full time-dependent differential Master Equation. Instead one can simply find the solution to the much simpler set of linear algebraic equations satisfied by $T(S_f|S_i)$.

For continuous variables, the MFPT obeys the corresponding Backward Fokker-Planck equation \cite{chou-review-2014,redner-book,gardiner-book-2003,jacobs2010stochastic},
\begin{align}
 A(s_{i})\pd_{s_{i}}T(s_f|s_i)+ \frac{1}{2}B(s_{i})\pd^{2}_{s_{i}}T(s_f|s_i)=-1.
\end{align}
\emph{Heuristics.} One can examine the simple logic behind the cumbersome mathematics of the Backward Equations with the following simple example. Consider a symmetric and homogeneous random walker on a lattice, hopping with equal rates, $r$, to the left or right, thus changing its position, $x$, by $\pm a$. We wish to find the Mean First Passage time, $T(x_{0})$, of the random walker arriving at $x=0$, starting from some position $x_0$. Following the above ``backward" arguments, any trajectory from $x_0$ to $0$ can be decomposed into two mutually exclusive families of paths:  one consisting of first jumping to the left, i.e.,  to $x_0-a$ and then going to $x=0$ from there, and the other in which the random walker first jumps to the right, to $x_0+a$, and then proceeds to $x=0$. For either family the first step, being a random Poisson process, takes a time $1/r$ on average. Thus the MFPT is found by averaging over these two families of equiprobable trajectories:
\begin{equation}\label{eq-mfptexample}
T(x_0) =\frac{1}{2}\le(\frac{1}{r} + T(x_0-a)\ri)+\frac{1}{2}\le(\frac{1}{r} + T(x_0+a)\ri).
\end{equation}\
Rearranging terms,
\begin{equation}
\frac{r}{2}T(x_0-a)+\frac{r}{2}T(x_0+a)-rT(x_0)=-1,
\end{equation}
in agreement with equation (\ref{eq-MFPT-backward}) above. Taking the limit $a\rightarrow 0$, we recover the equation satisfied by $T(x_{0})$ in terms of the continuous case Backward Fokker-Planck operator:
\begin{equation}
D\frac{\partial^2 }{\partial x_0^2}T(x_0) =-1.
\end{equation}

\subsubsection{First Passage Processes with multiple absorbing states.}\label{sec-framework-multiple-exits}
More complicated First Passage Processes can be addressed within the same framework. A question that arises frequently is the following. What is the FPT distribution for reaching the state $S_f$ for the first time, \emph{without} passing through a set of other states $\{S_{f'}\}$ before that? The answer to this question can be obtained following a similar prescription as above, by considering an auxiliary problem with absorbing boundary conditions at $S_f$ \emph{as well as} the states $\{S_{f'}\}$. The main difference from the previous case is that the probability of reaching the final state $S_f$ is not equal to one anymore: some trajectories get to one of the states $\{S_{f'}\}$ first and should not be counted amongst the first passage trajectories to $S_f$.
The probability of reaching $S_f$ at time $t$, before any of the states $\{S_{f'}\}$, starting from the state $S_i$ at time $t=0$, is
\begin{align}
\mc{P}(S_f|S_i)=\int_{0}^{\infty}dt J(t, S_f|S_i),
\end{align}
where $J(t, S_f|S_i)$ is the probability flux into the state $S_f$ at time $t$. Noting that the probability of jumping directly from the state $S_i$ to any other state $S$ is $q(S,S_i)=\frac{\alpha(S,S_i)}{\sum_{S'}\alpha(S',S_i)}$, and using the backward reasoning, we get,
\begin{align}
\mathcal{P}(S_f|S_i)=q(S_f,S_i)+\sum_{S'\neq S_f,S_{f'}}q(S',S_i)\mathcal{P}(S_f|S').
\end{align}
The first term is the probability to go directly to $S_f$ from $S_i$ and the second is the probability to first go to some state $S'$ and then go to $S_f$ from there (without passing through any of the states $S_{f'}$). In other words, the vector $\boldsymbol{\mathcal{P}}$, whose $i$-th component is $\boldsymbol{\mathcal{P}}(S_f|S_i)$, satisfies the equation
\begin{align}
\mc{M}_b\cdot\boldsymbol{\mathcal{P}}=-\boldsymbol{\mathcal{V}},
\end{align}
where $\boldsymbol{\mathcal{V}}$ is a vector with components $\boldsymbol{\mathcal{V}}_i=\alpha(S_f,S_i)$.

Now, the normalized probability distribution of the First Passage Times into the state $S_f$ is given by
\begin{equation}
F(t,S_f|S_i)=J(t, S_f|S_i)/\mc{P}(S_f|S_i).
\end{equation}
Using arguments similar to those leading to Eq.~\eqref{eq-mfptexample}, the Mean First Passage time can be shown to satisfy the following equation:
\begin{equation}\label{eq-backward-directional-time}
\sum_{S'}(\mc{M}_b)_{S_i S'}(\mc{P}(S_f|S')T(S_f|S')) =-\mc{P}(S_f|S_i).
\end{equation}
For continuous variables, the corresponding Fokker-Planck equation is
\begin{equation}\label{eq-backward-directional-prob-FP}
A(s_i)\pd_{s_i} \mc{P}(s_f|s_i)+\frac{1}{2}B(s_i)\pd_{s_i^2}\mc{P}(s_f|s_i)=0,
\end{equation}
with the boundary conditions $\mc{P}(s_f|s_f)=1$ and $\mc{P}(s_{f}|s_{f'})=0$ for $f'\neq f$. For the MFPT,
\begin{equation}\label{eq-backward-directional-time-FP}
A(s_i)\pd_{s_i} (T(s_f|s_i)\mc{P}(s_f|s_i)+\frac{1}{2}B(s_i)\pd_{s_i^2}(T(s_f|s_i)\mc{P}(s_f|s_i)=-1,
\end{equation}
with the boundary conditions $(T(s_f|s_{f'})\mc{P}(s_f|s_{f'}))=0$ for all $f'$, including $f=f'$.
The applications of these formal expressions are illustrated below.

\subsubsection{Kramers' method.}\label{sec-kramers}
Another method, originally used by Kramers in the famous 1940 paper \cite{kramers1940}, can be used for the calculation of the Mean First Passage Times and probabilities.  In order to calculate the MFPT from a state $S_i$ to a state $S_f$, Kramers considered the auxiliary problem with an absorbing condition at $S_{f}$ and a constant flux $J$ entering at the state $S_i$. The mean time,  $T_K$, that the particles spend in the system, traveling from the state $S_i$ to the state $S_f$ can be calculated from the average occupancies of all states, $N(S,t)$, which obey the same Master Equation as the probability distributions of the individual particles, Eq.~\eqref{eq-masterequation0} with the extra flux term.  Intuitively, in steady state, the flux through the system obeys the following relation
\cite{Hanggi2009}:
\begin{align}\label{eq-TK}
J=\sum_{S\neq S_{f}} N(S)/T_K.
\end{align}
It can be rigorously shown that the Kramers' time $T_K$ is identical to the actual MFPT from $S_i$ to $S_f$, proven by by Reimann, Schmid and H\"{a}nngi in \cite{Reimann1999} (see also \cite{Hanggi2009}).

\emph{Formal derivation.}
At steady state, the vector of occupancies, $\boldsymbol{N}$, satisfies the equation $\pd_t\boldsymbol{N}=\mc{M}_f\cdot\boldsymbol{N}+\boldsymbol{J}=0$, where $J_{S}=J\delta_{S,S_i}$. Thus,  $N(S)=-J(\mc{M}_f^{-1})_{SS_i}$ and
\begin{align}
T_K=\sum_S N(S)/J=-\sum_{S}(\mc{M}_f^{-1})_{SS_i},
\end{align}
from Eq. \ref{eq-TK}. On the other hand, the vector of the MFPT's, $\boldsymbol{T}$, is the solution of the Backward Master Equation Eq.(\ref{eq-backward-master-operator-form}), $\mc{M}_b\cdot\boldsymbol{T}=-\boldsymbol{I}$, where $\boldsymbol{I}$ is the unity vector with components $I_S=1$ for all $S$. In other words,
\begin{equation}
T(S_f|S_i)=-\sum_{S}(\mc{M}_b^{-1})_{S_iS}=-\sum_{S}(\mc{M}_f^{-1})_{SS_i}=T_K,
\end{equation}
where we have used the fact that the forward operator $\mc{M}_f$ is the transpose of the backward operator: $(\mc{M}_f)_{SS'}=(\mc{M}_b)_{S'S}$.

Kramers' method can be also extended to  calculation of the probabilities, but not the times of exit into multiple absorbing states. For instance, the probability to exit through state $S_f$ starting from state $S_i$ is
\begin{align}
\mc{P}(S_f|S_i)=J(S_f)/J,
\end{align}
where $J(S_f)$ is the steady state flux into the state $S_f$ \cite{zilman-plos-2007}.
Although less general and non-generalizable to finding the probability distributions, Kramers' method is often a useful and convenient way to calculate Mean First Passage Times and probabilities.

\section{Applications.}
\subsection{Channel transport. }
\subsubsection{Background.}
Ubiquitous channels and transporters shuttle various materials into and out of the cell, as well as between different cellular compartments. Examples include porins in bacteria, nuclear pore complex in eukaryotic cells, transport of polypeptides into the endoplasmic reticulum, ion channels and many others. Their functioning provides inspiration for the creation of bio-mimetic nano-transporters for technological applications. During the past decade, research of transport through biological and bio-mimetic transporters has seen increased application of precise and quantitative biophysical techniques that allow the resolution of the durations of the single molecule transport events on the single channel level, in parallel with the development of the appropriate mathematical analysis tools. Combination of the experimental and theoretical work has resulted in the development of a conceptual framework for the explanation of the transport specificity and efficiency of such nanochannels \cite{stein-book-2012,winterhalter-review-2008,kasianowicz-2001-protein-pore,bezrukov-kullman-maltoporin-2000,storm2-dekker-DNA-2005,kowalczyk-NatureN-2011,Goyal2010,Firnkes2010,rabin-dekker-2013,stewart-review-2007,musser-weidong-2000,Tu2013,weidong-3d-2013,grunwald2010vivo,berezhkovskii-times-2003,chou-zeolites-PRL-1999,kolomeisky-kotsev-2008,zilman-plos-2007}.

\begin{figure}[h]
\begin{center}
\resizebox{10 cm}{!}{\includegraphics{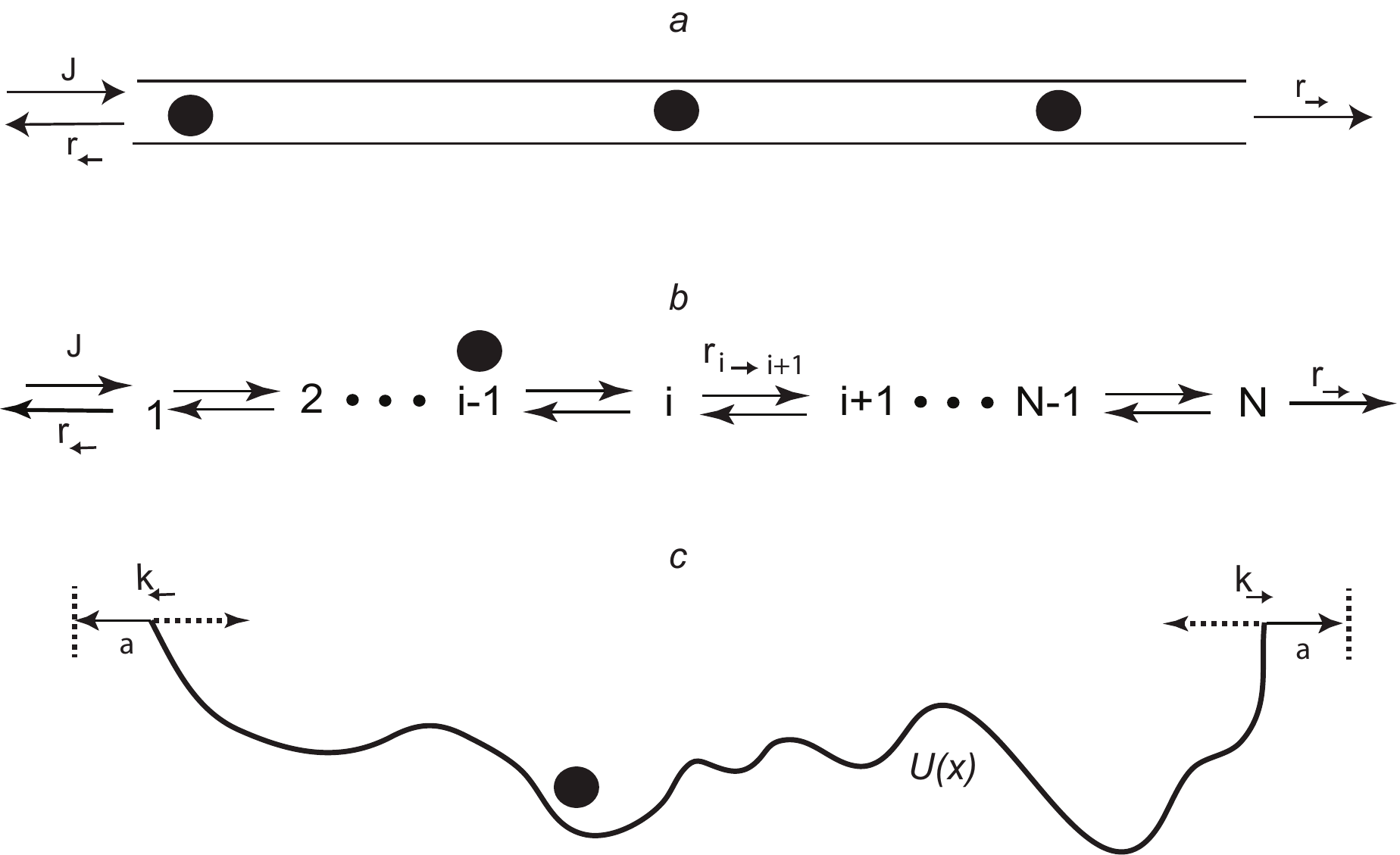}}
\caption{\textbf{Channel transport representations.} a) Schematic illustration of the channel transport. b) Discrete representation of the channel as a sequence of discrete sites. c) Particle movement in the channel is viewed as a continuous diffusion in an effective potential $U(x)$. The exit probabilities at the channel ends can be represented either via a radiation boundary condition or as absorbing boundary conditions located at a short distance $a$ (of the order of the particle size) from the channel ends (See text). All these models approximate transport as one-dimensional. Nevertheless, the FP methods can be extended to take into account the full three-dimensional nature of the channel transport \cite{licata-sticky-2009}.}
\label{fig-channel-representations}
\end{center}
\end{figure}

Mathematically, transport through a channel can be viewed as a First Passage process whose starting point is the entrance of the particle into the channel and its final point is the particle exit from the channel.  Figure \ref{fig-channel-representations} illustrates the different representations of the channel transport problem, discussed below. 
\subsubsection{Discrete channel representation: forward Master Equation method}\label{sec-discrete-channel}
Transport of a particle through a channel can be viewed as the hopping between discrete sites, as illustrated in Fig. \ref{fig-channel-representations}. The hopping rates can reflect the energetics, the external forces, local geometry or any other factors that affect the particle motion in the channel. The model itself is much more general than just a description of the channel transport. With the appropriate choice of rates, it has been used to describe molecular motors walking on a microtubule, DNA polymerase during transcription, RNA transcript moving through the ribosome during translation, or a transcription factor search of the binding site on the DNA \cite{kolomeisky-fisher-2007-motors-review,Slutsky2004,Kolomeisky2011,Chou2011,Hanggi2009,Bressloff-Review-2014}.

The particles start at site $i=1$ and hop inside the channel between the adjacent sites with the rates $r_{i\rightarrow i\pm 1}$ until they either translocate through the channel, exiting from  site $N$ with the rate $r_{\rightarrow}$, or exit the the "wrong side" - site $1$, with the rate $r_{\leftarrow}$ \cite{Kolomeisky2007,Berezhkovskii2007,bezrukov-sites-2005,zilman-BJ-2009,zilman-pearson-bel-PRL-2009}. The probability $P_i(t)$ for the particle to be at site  $i$ at time $t$ then obeys the (forward) Master Equation
\begin{align}
\pd_{t}P_i(t)&=r_{i-1\rightarrow i}P_{i-1}(t)+r_{i+1\rightarrow i}P_{i+1}(t)-(r_{i\rightarrow i+1}+r_{i\rightarrow i-1})P_i(t)\;\text{for}\;1<i<N\nn\\
\pd_{t}P_1(t)&=r_{2\rightarrow 1}P_{2}(t)-r_{\leftarrow}P_{1}(t)\;,\;\pd_{t}P_N(t)=r_{N-1\rightarrow N}P_{N-1}(t)-r_{\rightarrow}P_{N}(t).
\end{align}
In the matrix form,
\begin{align}\label{eq-forward-master-discrete-channel}
\pd_{t}\textbf{P}(t)=\hat{M}\cdot\textbf{P}(t),
\end{align}
where the vector $\textbf{P}(t)=(P_1(t),..,P_i(t),..,P_N(t))$ and the tri-diagonal matrix $\hat{M}$ has the following elements: $\hat{M}_{i,j}=r_{j\rightarrow i}\delta_{j,i\pm 1}-(r_{j\rightarrow j+1}+r_{j\rightarrow j-1})$ for $1<j<N$, $\hat{M}_{1,1}=-(r_{\leftarrow}+r_{1\rightarrow 2}),\hat{M}_{N,N}=-(r_{\rightarrow}+r_{N\rightarrow N-1})$.

For a particle starting at site $i=1$, the initial condition is $P_i(0)=\delta_{i,1}$, and the solution to Eq. (\ref{eq-forward-master-discrete-channel}) is $P_i(t)=\left(e^{\hat{M}t}\textbf{P}(0)\right)_i=\left(e^{\hat{M}t}\right)_{1,i}$. According to Section \ref{sec-framework-multiple-exits}, the probability to translocate through the channel, exiting through  site $N$, is  the integral of the probability flux out of site $N$:
\begin{align}
\mc{P}_{\rightarrow}=\int_0^{\infty}r_{\rightarrow}\left(e^{\hat{M}t}\right)_{1,N}dt= r_{\rightarrow}\left(\hat{M}^{-1}\right)_{1,N}.
\end{align}
The probability density of the transport times distribution is then
$$F_{\rightarrow}(t)=r_{\rightarrow}\left(e^{\hat{M}t}\right)_{1,N}/\mc{P}_{\rightarrow}$$
and the Mean First Passage Time is
\begin{align}\label{eq-MFPT-matrix-method}
T_{\rightarrow}=\frac{1}{\mc{P}_{\rightarrow}}\int_0^{\infty}t F(t)dt=\frac{\hat{M}^{-2}_{1,N}}{\hat{M}^{-1}_{1,N}}.
\end{align}

\emph{Special case: uniform and symmetric channel.} For a uniform and symmetric channel, where all the internal rates are equal, $r_{i\rightarrow i\pm 1}=r$, and the exit rates at the ends are equal to each other, $r_{\rightarrow}=r_{\leftarrow}=r_o$, the transport probability and the time can be calculated analytically \cite{zilman-pearson-bel-PRL-2009}:
\begin{align}\label{eq-Pf-discrete}
\mc{P}_{\rightarrow}=\frac{1}{2+(N-1)r_o/r}\;\;\;\text{and}\;\;\; T_{\rightarrow}\simeq \frac{N}{6r_o}(6+6Nr_0/r+(Nr_o/r)^2).
\end{align}

This equation has interesting physical consequences. In the diffusion dominated regime, $Nr_o/r\gg 1$, the probability of translocating is small: $\mc{P}_{\rightarrow}\simeq\frac{1}{Nr_o/r}\ll 1$, because most of the particles exit from site $1$ soon after the entry, without translocating. In this regime, the transport time displays the familiar scaling with the channel length: $T_{\rightarrow}\simeq N^2/r$.

By contrast, in the opposite regime, $Nr_o/r\ll 1$, which corresponds to trapping the particle in the channel, the rate-limiting step is the exit from the channel end. In this case,  the transport time scales linearly with the channel length: $T_{\rightarrow}\simeq N/r_o$, illustrating the often non-intuitive behavior of the First Passage Times. Despite the fact that the transport time is long in this limit, the transport  probability increases to $\mc{P}_{\rightarrow}=1/2 $ independent of the parameters. This counter-intuitive  fact was first realized in the context of the facilitation of oxygen transport in tissue by myoglobin \cite{wyman-myoglobin-facilitated-1966}. More recently, facilitation of channel transport by molecular trapping, corresponding to small $r_o/r$, has emerged as the explanation of the specificity of channel transport (see also the next section) \cite{bezrukov-ptr-2002,zilman-plos-2007,kolomeisky-kotsev-2008,bezrukov-antibiotics-PNAS-2002,bezrukov-kullman-maltoporin-2000}. 

The total mean residence time in the channel, averaged over both translocating and returning particles,  is $T_{tot}=\mc{P}_{\rightarrow}T_{\rightarrow}+\mc{P}_{\leftarrow}T_{\leftarrow}=\frac{N}{2r_o}$. Note that it scales linearly with the channel length, counter to our intuition about the diffusion times.

In principle, the Mean First Passage Times can be calculated explicitly for any set of transition rates either using Eq. (\ref{eq-forward-master-discrete-channel}) and calculating the probability flux, or by solving the Backward Master Equation. The final answer is obtained in terms of large  combinations of the transition rates and is very cumbersome. These transport times and probabilities can also be obtained using the Kramers method (Section \ref{sec-kramers} and \cite{zilman-plos-2007,zilman-bel-jphys-2010}). The methods of this section can also be used for the calculation of FPT distributions \cite{zilman-bel-jphys-2010,chou-review-2014,chou-nowak-virus-2009}. The reader is referred to \cite{mfpt-random-2003-caceres,karlin-book-1998,Mirny2009} for details; see also Section \ref{sec-wofsy} below.

\subsubsection{Continuous coordinate representation: backward Fokker-Planck approach.}
Particle motion in the channel can also be represented as continuous diffusion in a potential $U(x)$ with the diffusion coefficient $D(x)$, which, in principle, can be spatially dependent. The discrete and the continuous models can be connected by relating the hopping rates between adjacent sites, $r_{i\rightarrow i\pm 1}$,  to the energy differences: $r_{i\pm 1\rightarrow i}=2D/d^2e^{-(U_i-U_{i\pm 1})/2kT}$, where $d$ is the inter-site distance. However, any choice of rates that satisfies the detailed balance condition,  $r_{j\rightarrow i}/r_{i\rightarrow j}=e^{-(U_i-U_j)/kT}$, is physically acceptable. In the continuous representation, the exit rates from the channel at $x=0$ and $x=L$ can be taken into account using the radiative boundary conditions at the channel ends: $\pd_{x}(x,t)|_0=\frac{1}{k_{\leftarrow}}p(0,t)$ and  $\pd_{x}p(x,t)|_L=-\frac{1}{k_{\rightarrow}}p(L,t)$ \cite{bezrukov-ptr-2002,muthukumar-book}. The constants $k_{\leftarrow}, k_{\rightarrow}$ determine the probability of the actually exiting the channel once it reaches the boundary, or getting ``reflected'' back inside. A completely absorbing boundary corresponds to $k=0$, while $k=\infty$ corresponds to a completely reflective boundary. Thus, they can be related to the rates, $r_{\rightarrow}$ and $r_{\leftarrow}$, of the discrete case that also reflect the probabilities of the particle at the exit site to leave the channel, $\frac{r_{\leftarrow}}{r+r_{\leftarrow}}$ and $\frac{r_{\rightarrow}}{r+r_{\rightarrow}}$.

According to Section \ref{sec-framework-multiple-exits}, the translocation probability $\mc{P}_{\rightarrow}(x)$ to exit through $x=L$, starting from an arbitrary position $x$, satisfies the stationary Backward Fokker-Planck equation (\ref{eq-backward-directional-prob-FP}),  compactly written as
\begin{equation}
\frac{\partial }{\partial x}\le(D(x)e^{-U(x)/kT}\frac{\partial}{\partial x}\le(e^{U(x)/kT}\mc{P}_{\rightarrow}(x)\ri)\ri)=0,
\end{equation}
with  the boundary conditions $\pd_{x}\mc{P}_{\rightarrow}(x)|_0=\frac{1}{k_{\leftarrow}}\mc{P}_{\rightarrow}(0)$ and  $\pd_{x}\mc{P}_{\rightarrow}(x)|_L=-\frac{1}{k_{\rightarrow}}\mc{P}_{\rightarrow}(L)$ \cite{bezrukov-ptr-2002,berezhkovskii-times-2003,Berezhkovskii1999}.

The directional transport times $T_{\rightarrow}(x)$ can be calculated from  the corresponding backward equation   (\ref{eq-backward-directional-time-FP}),
\begin{equation}
\frac{\partial }{\partial x}\le(D(x)e^{-U(x)/kT}\frac{\partial}{\partial x}\le(e^{U(x)/kT}T_{\rightarrow}(x)\mc{P}_{\rightarrow}(x)\ri)\ri)=-\mc{P}_{\rightarrow}(x),
\end{equation}
with  the boundary conditions $\pd_{x}(T_{\rightarrow}(x)\mc{P}_{\rightarrow}(x))|_0=\frac{1}{k_{\leftarrow}}(T_{\rightarrow}(0)\mc{P}_{\rightarrow}(0))$
and\newline $\pd_{x}(T_{\rightarrow}(x)\mc{P}_{\rightarrow}(x))|_L=-\frac{1}{k_{\rightarrow}}(T_{\rightarrow}(L)\mc{P}_{\rightarrow}(L))$.

For $k_{\rightarrow}=k_{\leftarrow}$ and $U(0)=U(L)$, the above equations give for the transport probability, $P_{\rightarrow}\equiv \mc{P}_{\rightarrow}(0)$,
\begin{equation}
\mc{P}_{\rightarrow}=\frac{1+k\int_0^L dye^{U(y)/kT}/D(y)}{2+k\int_0^L dye^{U(y)/kT}/D(y)},
\end{equation}
and for the transport time $T_{\rightarrow}\equiv T_{\rightarrow}(0)$,
\begin{equation}
T_{\rightarrow}=\frac{1}{k}\mc{P}_{\rightarrow}\le(\int_0^L\le[1+\int_0^x\frac{e^{U(y)/kT}}{D(y)}dy\ri]\le[1+\int_x^L\frac{e^{U(y)/kT}}{D(y)}dy\ri]e^{-U(x)/kT}dx\ri).
\end{equation}

\emph{Special case: uniform channel.} For a uniform potential profile and constant diffusion coefficient,  $U(x)=E$ and $D(x)=D$ for all $x$, one gets for the transport probability
\begin{equation}
\mc{P}_{\rightarrow}=\frac{1}{2+\frac{k}{D}L e^{E/kT}},
\end{equation}
and time,
\begin{equation}
T_{\rightarrow}=\frac{L}{6k}e^{-E}\frac{6+6\frac{L k}{D}e^{E}+(\frac{L k}{D})^2e^{2E}}{2+\frac{Lk}{D}e^{E}}.
\end{equation}
With the appropriate identification of $k$, these expressions become identical to  the discrete channel model, Eq. (\ref{eq-Pf-discrete}). In the context of channel transport, one is typically interested in molecular trapping inside the channels, $E<0$. In particular, in the limit of short channel and strong trapping,  $Lke^{E}/D\ll 1$, the translocation time $T_{\rightarrow}\simeq Le^{-E}/k$ is proportional to the channel length and exponentially increases with the trapping energy $|E|$. Conversely, the limit of long channels, $Lke^{E}/De\gg 1$, the transport is dominated by diffusion and the transport time obeys the familiar scaling with the channel length $L$, $T_{\rightarrow}\simeq  L^2/D$ .

\emph{Physical choices of the exit rates and the radiative constants: equivalence of different models.}
The choice of the exit rates in the discrete site method and the $k$'s in the radiation boundary method depends on the physical problem under consideration. For channel transport, they can be determined from the coupling of the quasi one-dimensional diffusion inside the channel to the three dimensional diffusion outside. This can be performed  either in the forward \cite{zilman-BJ-2009} or the backward  \cite{bezrukov-fluctuations-2000} formalism and results in $k=\frac{4D_o}{\pi a}$ in the radiation boundary condition method and $r_o/r=\frac{D_o}{D}\frac{L}{a}e^{E}$ in the discrete site method; $a$ is the channel radius and $D_o$ is the diffusion coefficient outside the channel. Finally, identifying $r=2D/d^2$ ($d$ is the inter-site distance), the expressions for the transport probabilities and times obtained by the discrete and the continuous methods become equivalent up to a numerical factor of $4/\pi$ in the denominator (Eq. (\ref{eq-Pf-discrete})).

\subsubsection{Mapping onto one-dimensional diffusion.}\label{section-channel-simple-diffusion}
Another rendering of the channel transport, which approximates the transport also outside the channel as one-dimensional diffusion, is useful for the analysis of transport events through individual pores on the single molecule level \cite{Tu2013,kowalczyk-NatureN-2011,grunwald2010vivo,musser-weidong-2000,zilman-plos-2007}. In this representation, the particle starts from the position $x=0$ (channel entrance) and performs one-dimensional diffusion in the potential $U(x)$ until it reaches an absorbing boundary at either $x=-a$ or $x=L+a$, corresponding to the exit from the channel.

As discussed in Section \ref{sec-framework-multiple-exits}, the transport probability $\mc{P}_{\rightarrow}(x)$ to reach $L+a$ starting from $x$ satisfies the stationary Backward Fokker-Planck equation  with the  boundary conditions $\mc{P}_{\rightarrow}(-a)=0$ and $\mc{P}_{\rightarrow}(L+a)=1$ \cite{gardiner-book-2003,jacobs2010stochastic,zilman-plos-2007},
\begin{align}
D\frac{\partial }{\partial x}\le(e^{-U(x)/kT}\frac{\partial}{\partial x}\le(e^{U(x)/kT}\mc{P}_{\rightarrow}(x)\ri)\ri)=0,
\end{align}
which gives
\begin{align}
\mc{P}_{\rightarrow}(x)=\frac{\int_{-a}^xe^{U(y)/kT}dy}{\int_{-a}^{L+a}e^{U(y)/kT}dy}.
\end{align}
Note that $\mc{P}_{\rightarrow}(x)$  is independent of the diffusion coefficient $D$.

For a flat potential profile, $U(x)/kT=E$  for $0<x<L$ (inside the channel) and $U(x)=0$ outside the channel, assuming that the diffusion coefficient is the same inside and outside the channel, the transport probability $\mc{P}_{\rightarrow}\equiv \mc{P}_{\rightarrow}(0)$ becomes
\begin{equation}
\mc{P}_{\rightarrow}=\frac{1}{2+\frac{L}{a}e^{E/kT}}.
\end{equation}
Note that it is equivalent to the expression obtained in the discrete channel representation.

The mean transport time obeys the corresponding backward Fokker-Planck equation
\begin{align}
D\frac{\partial }{\partial x}\le(e^{-U(x)/kT}\frac{\partial}{\partial x}\le(e^{U(x)/kT}\mc{P}_{\rightarrow}(x)T_{\rightarrow}(x)\ri)\ri)=-\mc{P}_{\rightarrow}(x).
\end{align}
with the boundary conditions $\mc{P}_{\rightarrow}(L+a)T_{\rightarrow}(L+a)=\mc{P}_{\rightarrow}(-a)T_{\rightarrow}(-a)=0$. For the negative and flat potential profile, $U(x)/kT=E$, it yields

\begin{equation}
T_{\rightarrow}=\frac{aL}{2D}\le(e^{-E/kT}(1-a/L)+a/L\ri)\simeq \frac{aL}{2D}e^{-E/kT}\;\;\;\text{for}\;\;\; |E|/kT\gg 1.
\end{equation}

This expression for the forward time is  qualitatively similar to the expressions obtained using the discrete and the radiative boundary conditions methods. These results are also closely related to the transport of the long chains, such as flexible macromolecules, through small pores (see below).

\subsubsection{Multiple particles in the channel.}
Until now, we have considered a single particle in the channel. However, a channel can contain several particles simultaneously, which interfere with each other's movement. Description of the movement of an individual particle within the flux of other particles (known as the ``tracer'' particle) is a complicated problem because the motions of the neighboring particles are correlated. In this case, there is no closed Master Equation for the probability distribution of the tracer particle, and understanding single molecule transport in this regime remains a major challenge.

\emph{Exact solutions.}
It is possible to obtain some exact results for mean residence times even for channels with large numbers of particles  although the results are typically cumbersome \cite{Rodenbeck1997,karger-tracer-exchange-1997,schutz-exact-single-file-1997}. Here, we briefly sketch the main points of the derivation for the  case of single file transport in a uniform channel in equilibrium with a  solution of particles \cite{Rodenbeck1997}. Most generally, the system of multiple particles in a channel is described by the multi-particle probability function $P(\vec{x},t|\vec{y})$ that the vector of particles' positions is $\vec{x}$ at time $t$, starting from the initial vector $\vec{y}$ \cite{Barkai2010,chou-review-2014,Rodenbeck1997}.  The crucial insight is that because the particles cannot bypass each other, the initial order of the particles is conserved: if $y_m<y_n$ for any two particles  at the initial time, it implies that $x_m<x_n$ for all future times. That is, the parts of the phase space accessible to these particles  are bounded by the planes defined by the condition $x_n=x_m$ in the vector space $\vec{x}$. This implies a reflective boundary condition at the $x_m=x_n$ plane for any two different particles $m$ and $n$,
\begin{align}
\pd_{n}P(\vec{x},t|\vec{y})=0\;\;\text{and thus}\;\;\left(\pd_{x_m}P(\vec{x},t|\vec{y})-\pd_{x_n}P(\vec{x},t|\vec{y})\right)=0
\end{align}
where $\pd_{n}$ denotes derivative normal to the plane defined by $x_n=x_m$. One can then use the multi-dimensional generalization of the image method to compute $P(\vec{x},t|\vec{y})$ and  the survival probability of the ``tracer'' particle by integrating out all other coordinates \cite{Rodenbeck1997,redner-book,1999-jackson-zh}. For single-file transport, the ``tracer'' particle is known to perform anomalous diffusion with the mean square displacement varying with time as $\langle \Delta x^2\rangle\sim t^{1/2}$ instead of the familiar diffusion law $\langle \Delta x^2\rangle\sim t$. This type of motion can be treated within the anomalous diffusion formalism, which, in principle allows calculation of the appropriate First Passage times and probabilities \cite{Barkai2010,karger-hagn-anomalous-bservation-1996,barkai-metzler-anomalous-review-2014,klafter-anomalous-review}.

\emph{Mean field approximations.} Insights into the first passage times of interacting particles in crowded channels can be obtained using the mean field/effective medium approach that approximates the effect of the other particles on  the "tracer" particle by the average steady state density  (see \cite{kutner-1981,Chou2011,chou-review-2014,chou-nowak-dynamic-2007}). For the discrete hopping model of Section \ref{sec-discrete-channel}, in the mean field approximaton the problem reduces to the single particle case with appropriately modified hopping rates, $r_{i\rightarrow j}\rightarrow r_{i\rightarrow j} (1-\bar{n}_j)$, where $\bar{n}_i$ is the average steady state occupancy of site $i$ \cite{zilman-pearson-bel-PRL-2009,zilman-bel-jphys-2010}. Although this method neglects the correlations between the particles and misses many important properties of crowded diffusion, it gives reasonable approximations for the MFTP.

\begin{figure}[h]
\begin{center}
\resizebox{10 cm}{!}{\includegraphics{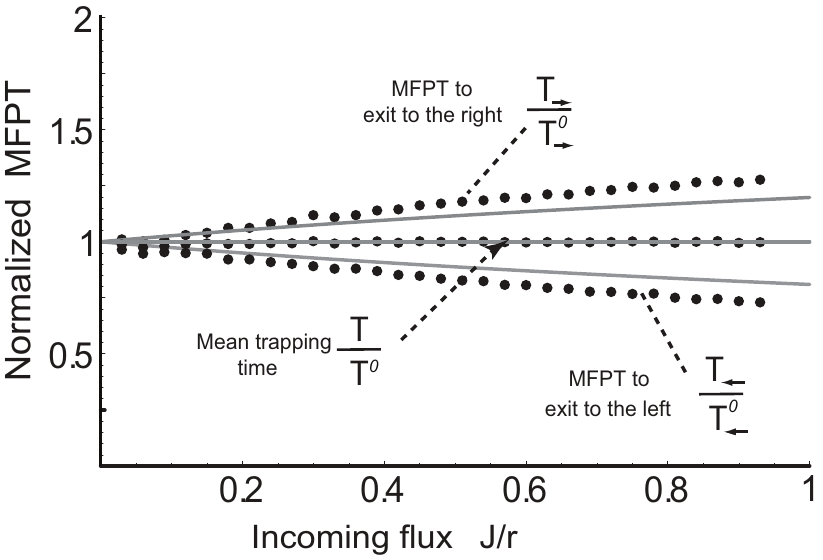}}
\caption{\textbf{First Passage Time of the ``tracer'' particle within steady state flux.} Mean translocation time $T_{\rightarrow}$ of an individual particle within a non-equilibrium steady state flux through the channel, normalized by the transport time in an empty channel, $T_{\rightarrow}^0$, as a function of the flux through the channel. The lines are analytical results; dots are the simulations. Based on \cite{zilman-pearson-bel-PRL-2009}.}
\label{fig-FPT-vs-J}
\end{center}
\end{figure}

For a uniform and symmetric channel with a steady state flux $J$ of particles impinging at the channel entrance, the dynamics of the ``tracer'' particle is then described by the discrete random walk model defined in Eq. (\ref{eq-forward-master-discrete-channel}), with the transition matrix
$\hat{M}_{i\pm 1,i}=r(1-\bar{n}_{i\pm 1})\;\;\hat{M}_{i,i}=-r(2-\bar{n}_{j+1}-\bar{n}_{j-1})$ and $0$ otherwise \cite{zilman-pearson-bel-PRL-2009}. The resulting analytical expressions  are cumbersome, and the outcomes are summarized in Fig. \ref{fig-FPT-vs-J}.

The main conclusion is that the crowding increases the average translocation time, while decreasing the average  time of abortive transport events, in which the particle returns from site $1$.  Surprisingly, for the uniform and symmetric process, the transport probability $P_{\rightarrow}$ and the overall residence time $T$ of the ``tracer'' particle are the same as in the single-particle case of Section \ref{sec-discrete-channel},
\begin{align}
\mc{P}_{\rightarrow}=\frac{1}{2+r_o(N-1)/r}\;,\;\;\;\;\;
T=\mc{P}_{\rightarrow}T_{\rightarrow}+\mc{P}_{\leftarrow}T_{\leftarrow}=\frac{N}{2r_o}.
\end{align}
This is a consequence of the cancellation of correlations for certain averaged quantities in interacting random walks on isotropic lattices \cite{kutner-1981}. 

\emph{Dense regime.} The situation simplifies again in the limit of very high densities, when essentially all available space is occupied by the particles. This occurs, for instance, for the transport of water through nanochannels, such as aquaporins or nanotubes \cite{aquaporin-review-2010,berezhkovskii-hummer-PRL-2002,winterhalter-review-2008}. In this case, the particles can enter and exit the channel only through large collective motions of the whole train of particles occupying the channel, whereby the lead particle leaves the channel concurrently with the entrance of a new particle from the rear. We denote the probability density of the time intervals $\tau$ between such  collective motions as $\psi(\tau)$.

The channel can be modeled as a chain of $M$ sites, each occupied by one particle at all times. The probability of a particle to be at a position $m$ along the channel at time $t$, starting from  $m_0$, obeys the following Master Equation,
\begin{align}
P(m,t|m_0)&=\delta_{m,m_0}(1-\int_0^td\t\psi(\t)) \nn \\
&+\frac{1}{2}\int_0^td\t\psi(\t)[P(m,t-\t|m_0+1)+P(m,t-\t|m_0-1)],
\end{align}
with the initial condition $P(m,0|m_0)=\delta_{m,m_0}$ and the boundary conditions \newline $P(0,t|m_0)=P(M+1,t|m_0)=0$, corresponding to the particle exiting the channel. If the time intervals between  large scale motions obey Poisson statistics with the mean inter-event time $1/k$, $\psi(t)=ke^{-kt}$, it can be shown using Laplace Transform that the above equation reduces to the familiar random walk on the discrete lattice \cite{berezhkovskii-hummer-PRL-2002}:
\begin{align}
\pd_{t}P(m,t|m_0)=\frac{k}{2}[P(m,t|m_0+1)+P(m,t|m_0-1)-2P(m,t|m_0)].
\end{align}
This allows the calculation of the transport times and probabilities using the methods described in Sections \ref{sec-discrete-channel} and \ref{section-channel-simple-diffusion}  \cite{berezhkovskii-hummer-PRL-2002}:
\begin{align}
\mc{P}_{\rightarrow}=\frac{1}{M+1}\;,\;\;\;T_{\rightarrow}=\frac{M(M+3)}{3k}.
\end{align}
Note that the transport probability is again the same as in the non-interacting particle case but the translocation time scales as $M^2$. The model also allows to calculate the times of more complicated collective motions, such as the interval between the exit times of the first and the last molecule of the train. This simple model of collective excitations in a strongly interacting system is in a very good agreement with the atomistic simulations \cite{berezhkovskii-hummer-PRL-2002}.

\subsubsection{Translocation of long chains through channels.}
The First Passage problem also arises in the context of translocation of long chains, such as DNA, RNA and unfolded proteins - and polymers in general - through nanopores. The biological examples include  translocation of unfolded proteins into the periplasm in bacteria and endoplasmic reticulum in eukaryotes. Research on the subject has been driven by the technological promise of such devices for DNA and RNA sequencing and protein sorting \cite{muthukumar-book,wanunu-DNA-review,stein-DNA-2013,storm2-dekker-DNA-2005,driessen-Sec-review-2011,wanunu-2014,golovchenko-2013}.

If the length $L$ of the polymer is much larger than the thickness of the pore, its motion can be viewed as the diffusion of the pore along the polymer, starting from $x=L$, not unlike the models of channel transport illustrated in Fig. \ref{fig-channel-representations}. Once the pore coordinate $x$ reaches zero, the polymer is considered to have translocated through the pore. In the simplest case, one can neglect the configurational entropy of the polymer outside the pore \cite{muthukumar-escape-2003,muthukumar-book}. Then the probability density of the pore being at a position $0<x<L$ along the polymer can be described by the Forward Fokker-Planck equation,
\begin{align}\label{eq-FP-polymer}
\pd_{t}p(x,t)=D(\pd_{x}^2p(x,t)-\frac{f}{kT}\pd_{x}p(x,t),
\end{align}
with the boundary conditions $p(0,t)=p(L,t)=0$, corresponding to the translocation and the return of the chain, respectively; $f<0$  is the external force (for instance, electric field) that pulls the polymer through the pore \cite{muthukumar-book}. The general solution of Eq. (\ref{eq-FP-polymer}) for the initial condition $x=x_0$ is
\begin{align}\label{eq-nelson-lubensky-solution}
p(x,t)=\frac{1}{L}\sum_{n=-\infty}^{\infty}e^{-w_nt}e^{\frac{(x-x_0)f}{2kT}}\sin(k_nx)\sin(k_nx_0),
\end{align}
where $k_n=\frac{\pi n}{L}$ and $w_n=D\left(k_n^2+\frac{1}{4}\left(\frac{f}{kT}\right)^2\right)$ \cite{lubensky-nelson-translocation-1999,redner-book}. Using the  Poisson identity, $\sum_{n=-\infty}^{\infty}f(n)=\sum_{m=-\infty}^{\infty}\tilde{f}(2\pi m)$, where $\tilde{f}(2\pi m)=\int dn f(n)e^{i2\pi mn}$,  Eq. (\ref{eq-nelson-lubensky-solution}) becomes
\begin{align}
p(x,t)=\frac{1}{\sqrt{4\pi Dt}}e^{\frac{(x-x_0)f}{2kT}}e^{-\frac{Dt}{4}(\frac{f}{kT})^2}\sum_{m=-\infty}^{\infty}\left[e^{-\frac{(x-x_0+2Lm)^2}{4Dt}}-e^{-\frac{(x+x_0+2Lm)^2}{4Dt}}\right],
\end{align}
which can be rewritten as
\begin{align}\label{eq-image-series}
p(x,t)=\frac{1}{\sqrt{4\pi Dt}}\sum_{m=-\infty}^{\infty}e^{\frac{fL}{kT}m}\left[e^{\frac{-(x-x_0-\frac{Df}{kT}t+2Lm)^2}{4Dt}}-e^{-\frac{(x+x_0-\frac{Df}{kT}t+2Lm)^2}{4Dt}e^{\frac{fx_0}{kT}}}\right].
\end{align}
Each term in these infinite series can be interpreted as an ``image'' particle with the starting point  at $-x_0$, $2L-x_0$, $-2L+x_0$, $2L+x_0$ etc., summed with the appropriate weights as to satisfy the boundary conditions $p(0,t)=p(L,t)=0$, analogous to the solutions to the Poisson equation in electrostatics \cite{weiss1983random,redner-book,1999-jackson-zh}.

The translocation probability can be calculated exactly:
\begin{align}
\mc{P}_{\rightarrow}=\int_0^{\infty}J(0,t)dt=\frac{1-e^{-\frac{f (L-x_0)}{kT}}}{1-e^{-\frac{fL}{kT}}},
\end{align}
where the probability flux into the absorbing boundary at $x=0$ is \mbox{$J(0,t)= |D\pd_{x}p(x,t)|_{x=0}$} (see Section \ref{sec-FPT-framework}). The normalized probability distribution of the translocation times, $F(t)$, and the average translocation time, $T_{\rightarrow}$ are
\begin{equation}
F(t)=J(0,t)/\mc{P}_{\rightarrow}\;,\;\;\;
T_{\rightarrow}=\lim_{x_0\rightarrow L}\left(\int_0^\infty tF(t)\right),
\end{equation}
which result in rather cumbersome expressions. However, the probability distribution of the translocation times can be approximated (for $L^2/(Dt)\gg 1)$ as
\begin{align}
F(t)\simeq \frac{2}{(Dt)^{3/2}}\left(\frac{L^2}{Dt}-1\right)e^{-(L-\frac{D|f|}{kT})^2/(4Dt)},
\end{align}
which has a maximum around $t_{\text{max}}=\frac{kTL}{D|f|}(1-5\frac{kT}{|f|L}+...)$ \cite{lubensky-nelson-translocation-1999}.  The maximum  of the probability density is an alternative characteristic of the typical translocation time. Note that for heavily asymmetric distributions, it can differ significantly from the mean time.

\emph{Approximations.}
For strong forces, or long channels, the typical translocation can be viewed as an almost deterministic motion in the direction of the force, with the mean ``velocity'' $v=\frac{D|f|}{kT}$. In this case, the probability that the chain does not translocate is low, and one can move the absorbing boundary condition at $x=L$ to $\infty$ \cite{muthukumar-book,Firnkes2010,Ling2013,li2013corrigendum}. This greatly simplifies the problem, which now has only one absorbing boundary condition at $x=0$. Taking the limit $L\rightarrow \infty$, Eqs. (\ref{eq-nelson-lubensky-solution}) and (\ref{eq-image-series}) reduce to \cite{redner-book,muthukumar-book}
\begin{equation}
p(x,t)=\frac{1}{\sqrt{4\pi Dt}}\left(e^{-\frac{\left(x-x_0-\frac{Dft}{kT}\right)^2}{4Dt}}-e^{-\frac{\left(x+x_0-\frac{Dft}{kT}\right)^2}{4Dt}}e^{\frac{x_0f}{kT}}\right).
\end{equation}
The distribution of the First Passage times is
\begin{align}
F(t)=\lim_{x_0\rightarrow L}J(0,t)=\frac{L}{t\sqrt{4\pi Dt}}e^{-\frac{-(L-\frac{D|f|t}{kT})^2}{4Dt}},
\end{align}
and mean translocation time is
\begin{align}
T_{\rightarrow}=\int_0^{\infty}\tau F(\tau)d\t=\frac{L^2}{D}\frac{kT}{|f|L}\frac{1+e^{-|f| L/kT}}{1-e^{-|f| L/kT}}
\end{align}
\cite{redner-book}. Note that to the first order in $\frac{kT}{|f|L}$, it is identical to $t_{\text{max}}$. As expected, for strong bias, $|f|L/kT\gg 1$, the translocation becomes an essentially deterministic motion with velocity $v=D|f|/kT$, so that $T_{\rightarrow}\simeq L/v\simeq t_{\text{max}}$ \cite{redner-book,slater-polymer-2008,stein-DNA-2013}.


\subsection{Receptor binding and adhesion.}
Another class of phenomena that are naturally described in the First Passage process formulation is the multivalent binding and adhesion - from macromolecular association to receptor signaling and viral cell entry \cite{lim-kapinos-2012-avidity,Tu2013,integrin-avidity-2003,chou-virus-entry-2007-true,rate-theories-for-biologists-2010,sethi-goldstein-multivalent-2011,gnana-binding-FPT-2013}.
In this section we review several recent works illustrating the applications of the First Passage methods to these problems.
\subsubsection{Viral particle binding and dissociation at the cell surface.}\label{sec-wofsy}
Typically, the first stage of viral entry into a target cell is the binding to the cell surface receptors.
The lifetime of a virus particle (virion) on the surface of a target cell is an important early determinant of the infection outcome. In the model of \cite{hlavacek1999dissociation}, the virion has $N$ sites on its surface that can bind receptors on the cell surface; the latter  are present in the surface concentration $C$. The virion is thus in one of the $N$ states: with $n=1,2,3,...N$ sites bound. The state with $n$ out of $N$ sites bound can transition into the state with $n-1$ bound sites, through breaking of one bond, with the rate $k_0$ for $n=1$ and $nk_{-1}$ for $n>1$. Alternatively, any of the unbound sites can form a new bond with a surface receptor,  resulting in a transition to the $n+1$ state, with the rate $(N-n)k_1C$. Physically, the rates $k_1$ and $k_{-1}$  reflect the local ``on'' and ``off'' rates, primarily determined by the binding energy, while $k_0$ reflects not only the the time of the local bond breaking but also the time of diffusing away from the cell surface. The process is illustrated in Fig. \ref{fig-virion-binding}. Note the similarity with the kinetic scheme for the particle in the channel of Fig. \ref{fig-channel-representations}. Similar models have been used to describe nanoparticle adhesion onto cell surface \cite{nanoparticle-decuzzi-2004adhesion}.

In the backward approach of Section \ref{sec-backward-equations}, the mean time to unbinding starting from $n$ bound sites, $T_n$, satisfies the equation
\begin{align}\label{eq-virion-wofsy}
T_n=\frac{1}{\mu_n+\lambda_n}+\frac{\lambda_n}{\lambda_n+\mu_n}T_{n-1}+\frac{\mu_n}{\lambda_n+\mu_n}T_{n+1},
\end{align}
where $\lambda_n=(N-n)k_1C$, $\mu_n=nk_{-1}$, $\mu_1=k_0$  \cite{hlavacek1999dissociation,hlavacek2002retention}.

Typically, the virion binding starts from just one site, $n=1$. In this case, the sequence of difference equations (\ref{eq-virion-wofsy}) can be solved analytically, giving for the average lifetime of the virion on the cell surface
\begin{align}\label{eq-diss-time-wofsy}
T_1=\frac{1}{k_0}\left[\frac{(1+KC)^N-1}{NKC}\right],
\end{align}
where $K=k_1/k_{-1}$ is the affinity of an individual site to a surface receptor \cite{hlavacek2002retention,karlin-book-1998}.

Physically, the binding affinity $K$ is the inverse of the dissociation constant $K_d$ and is related to the binding energy, $\epsilon>0$, as $K\sim e^{\epsilon/kT}$; the dissociation constant $K_d$ is sometimes colloquially referred to as the ``affinity'' as well. In the limit of weak binding or low surface receptor density, $KC\ll 1$, the dissociation time is $T_1\simeq 1/k_0$, indicating that the virion is most likely to escape immediately after binding without recruiting additional surface receptors. In the opposite limit of the very strong binding, $KC\gg 1$, $T_1\simeq \frac{(KC)^{N-1}}{N k_0}\simeq e^{N\epsilon/kT}$ and is exponential in the binding energy and the number of the binding sites, indicating that in this limit it is essentially cooperative binding that engages all the binding sites simultaneously.

\subsubsection{Multivalent binding: avidity}
Similar problems arise in many instances of multivalent binding. For example, transport proteins shuttling cargoes through the Nuclear Pore Complex (NPC) possess multiple binding sites to the hydrophobic residues located on the natively unfolded proteins located  within the Nuclear Pore Complex \cite{wente-rout-NPC-review-2010,stewart-review-2007}. First Passage theory can be used to analyze the results of single molecule fluorescence tracking experiments to infer the binding times and the the effective affinity of the transport factors to the hydrophobic repeats \cite{Tu2013}. Assuming that the hydrophobic repeats are present in volume concentration $F$ within the lumen of the NPC, the problem of the transport factor binding to the NPC becomes mathematically identical to the previous section. The dissociation time can be calculated from Eq. (\ref{eq-diss-time-wofsy}), which defines the effective ``off'' rate of the interaction. Together with an ``on'' rate of the first binding event, $k_{on}$, the effective dissociation rate determines the effective interaction affinity: $K_{\text{eff}}\equiv Nk_{on}\tau_{\text{off}}$. The factor $N$ arises because any one of the $N$ binding sites on the transport factors can bind a hydrophobic repeat first.

For a transport protein with four binding sites, expanding Eq. (\ref{eq-diss-time-wofsy}),
\begin{align}
K_{\text{eff}}=K_0(1+\frac{3}{2}KF+(KF)^2+\frac{1}{4}(KF)^3),
\end{align}
where $K_0=k_{on}/k_0$ is the affinity of the first binding event \cite{Tu2013}. This effective affinity $K_{\text{eff}}$
cannot be derived solely from the binding energies, but depends  on the concentration and the availability of the binding factors - effect known as the ``avidity'' for multi-valent interactions \cite{integrin-avidity-2003,lim-kapinos-2012-avidity,Tetenbaum-Novatt2012,sethi-goldstein-multivalent-2011}.

\subsubsection{Competition between viral dissociation, endocytosis and fusion.}\label{sec-chou}
Other possible outcomes of the virion binding to the cell surface, in addition to dissociation, were considered in \cite{chou-virus-entry-2007-true}. While the virion is bound to the surface receptors, it can fuse with the cell membrane and deliver its genetic material into the cell.  On the other hand, it can become engulfed by the cell membrane, endocytosed and targeted for destruction. The fates of the virus and of the infected cell are determined by which of these three processes completes first.
\begin{figure}[h]
\begin{center}
\resizebox{11 cm}{!}{\includegraphics{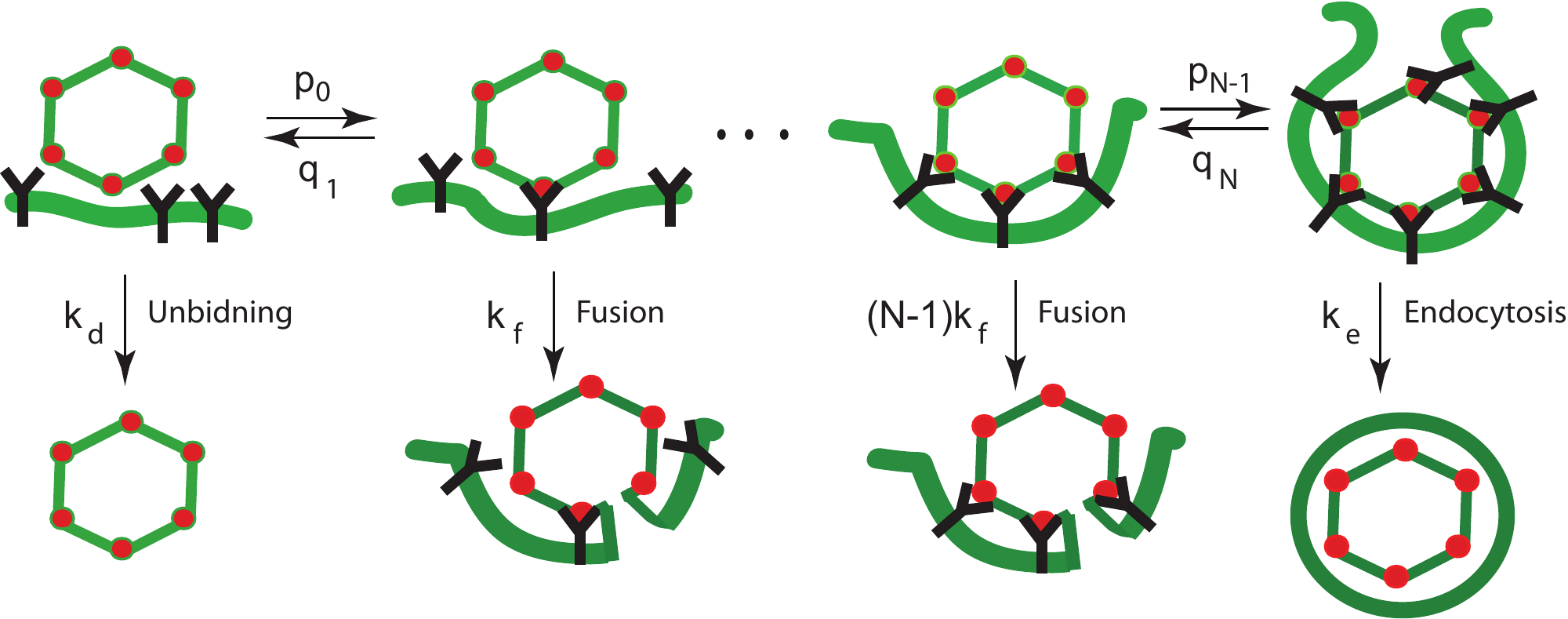}}
\caption{\textbf{Virus binding to the cell surface.} Schematic representation of the the kinetics of virus binding to the cell surface, reviewed in Sections \ref{sec-chou} and \ref{sec-wofsy}. In the model of Section \ref{sec-wofsy}, $k_e=k_f=0$.}
\label{fig-virion-binding}
\end{center}
\end{figure}
The kinetic scheme of the model is illustrated in Fig. \ref{fig-virion-binding}. The virion can be in one of the $N$ states with $n$ bound receptors plus the non-specifically adsorbed state at $n=0$, from which it can completely dissociate with rate $k_d$. In addition to the transitions from the state with $n$ bound sites to a state with $n\pm 1$ bound sites and the dissociation from state $n=0$, the process can terminate from $n=N$ through endocytosis with rate $k_e$, or via membrane fusion from any state $n$ with the rate $nk_f$. One  important difference of this model from Section \ref{sec-wofsy}, is that only binding sites that lie close to the circumference of the bound area of the virion can bind or unbind. Thus, the transition rates from state $n$ to $n-1$ (unbinding of one receptor) and to $n+1$ (binding an additional receptor), $q_n$ and $p_n$, respectively,  are  $p_n/k_1=q_n/k_{-1}\simeq \le(\frac{1-(1-2n/N)^2}{1-(1-2/N)^2}\ri)^{1/2}=(\frac{1}{2}N(1-(1-2n/N)^2)^{1/2}$ for $N\gg 1$. The $p_n$ and $q_n$ correspond to the $\lambda_n$ and the $\mu_n$ from Section \ref{sec-wofsy}. Overall, there is a race between the three possible outcomes: dissociation, endocytosis and fusion that occur with the corresponding probabilities,  $\mc{P}^d,\mc{P}^e,\mc{P}^f$ that sum to one: $\mc{P}^d+\mc{P}^e+\mc{P}^f=1$.

The Forward Master Equation for the probability to be in state $n$, $P_n(t)$ is
\begin{align}
\pd_{t}P_n(t)&=-(nk_f+p_n+q_n)P_n(t)+q_{n+1}P_{n+1}(t)+p_nP_{n-1}(t)\;\;\text{for}\;\;0<n<N\nonumber\\
\pd_{t}P_0(t)&=-(k_d+p_0)+q_1P_1(t)\\
\pd_{t}P_{N}(t)&=-(k_e+q_N)P_N(t)+p_{N-1}P_{N-1}(t).\nonumber
\end{align}
Using the numerical solutions of the Forward Master Equation, the probabilities of different outcomes can be calculated using $\mc{P}^f=\sum_nk_fn\int_0^{\infty}P_n(t)dt$, $\mc{P}^e=k_e\int_0^{\infty}P_N(t)dt$ and $\mc{P}^d=k_d\int_0^{\infty}P_0(t)$ \cite{chou-virus-entry-2007-true}.

These probabilities can also be calculated directly using the Backward Master Equation method. For instance, following Section \ref{sec-framework-multiple-exits}, the fusion probability $\mc{P}^f$ obeys the following set of equations:
\begin{align}
-nk_f&=-(nk_f+p_n+q_n)\mc{P}^f_n+p_n\mc{P}^f_{n+1}+q_n\mc{P}^f_{n-1}\;\;\text{for}\;\;n\geq 1\nn\\
0&=-(k_d+p_0)\mc{P}^f_0+p_0\mc{P}^f_1\\
-Nk_f&=-(Nk_f+k_e+q_N)\mc{P}^f_N+q_N\mc{P}^f_{N-1}\nn,
\end{align}
which can be solved by any method for solution of systems of linear algebraic equations.

Even this simplified model predicts rich behavior with  important biological implications. It be extended to include co-receptor binding and viral exocytosys from the infected cells \cite{holcman-viral-escape-2012,chou-nowak-virus-2009}.

\subsection{Single-cell growth and division.}
\subsubsection{Biological context.} The interplay between lengthscales and timescales in the context of cell growth and division can be cast as a First Passage Time problem by formulating how interdivision times are informed by the stochastic increase in cell size of individual cells. While the questions and modeling challenges in this context have been long appreciated~\cite{1949-monod-uq, 2001-koch-yg, 1991-cooper-vc,1991-cooper-sf, 1963-helmstetter, 1965-trucco-eq,1967-fredrickson-sr, 1968-donachie-mi, 1968-painter-gb,1982-trueba-ve, 1987-tyson-dt, 1985-tyson-yb, 1986-tyson-hs}, there is renewed interest~\cite{2011-scott-kx, 2012-amir-qv, -iyer-biswas-on, 2014-kennard-xh, 2014-pugatch-rg, 2010-scott-vn, Zhang2012, 2011-rading-kq, 2012-salman-cr, 2014-wolanski-bf, 2008-haeusser-bs, 2006-avery-ph, 2013-blaauwen-wl, 2003-matsumura-gn, 2012-velenich-vn, 2012-slusarczyk-ty, 2014-abner-vs, 2014-cohen-yu, 2014-scofield-ad, 2012-lan-oq, 2012-chien-mz, 2009-li-rc, 2009-wang-ys, Liang2010, 2006-cooper-rm, 2005-gitai-uq, 2007-goyal-kq, 2006-korobkova-ya, 2005-zhuravel-qf} due to the recent availability of large datasets for single cell growth trajectories and cell divisions, made possible by major breakthroughs in single cell technologies for unicellular organisms~\cite{2010-wang-cr, -iyer-biswas-kv, 2014-soifer-fh, 2014-kiviet-lj, 2007-talia-tg, 2013-santi-os, 2012-manalis-babymachine, 2013-ouyang-babymachine, 2012-brown-la, 2012-fritzsch-sy, 2011-mir-qf, 2012-moffitt-eg}.

\subsubsection{Formulating cell division as a first passage time problem.}
During each interdivision period, the size of a cell (assumed proportional to its mass ~\cite{2010-manalis-lognormal, 2010-godin-uq, 1991-cooper-vc}) increases according to ``the growth law'' on average ~\cite{2001-koch-yg, 1991-cooper-vc,1983-ingraham-qf}. Typically, this increase is either linear or exponential for unicellular organisms~\cite{1991-cooper-vc, 1983-ingraham-qf, 2014-amir-vn, -iyer-biswas-kv, 2010-wang-cr, 1968-kubitschek-lq}. There are three commonly considered scenarios for how the stochastically increasing cell size could inform the cell division. They are~\cite{1991-cooper-vc, 1983-ingraham-qf, 2014-amir-vn, -iyer-biswas-kv, 2014-osella-zr}, (i) ``absolute size thresholding''  (the ``sizer model''), in which the (stochastic) cell size attains a critical or threshold value at division; (ii) ``differential size thresholding''  (the ``adder model'') in which the thresholded variable is the change in cell size from its initial to final value; and, (iii) ``ratio size thresholding''  (the ``timer model'') in which the ratio of the size at division to the initial cell size is thresholded.  For clarity in elucidating the methodology, here we assume that the cells are in balanced (steady state) growth conditions, and that  intergenerational correlations are negligible. Together they imply that the statistics of growth and division are identical and independent for all generations of the cells.

Under these assumptions, the formulation of cell division as a first passage time problem requires that the following be specified: (i) a stochastic model for how cell size, $s$, increases with time, $t$, between divisions; (ii) the function of the cell size, $s$, that attains critical or threshold value, $\th$, at division; and (iii) appropriate initial conditions, including the initial distribution of cell sizes. While in all previous cases considered,  every member of the ensemble (i.e., each cell in the population) was assumed to be subjected to identical initial conditions, in this section we relax that condition to allow different cells to experience different initial conditions (such as different freshly divided cells having different sizes). This is an added source of stochasticity, namely, extrinsic noise in addition to the intrinsic fluctuations encoded in stochastic growth for  a given initial condition.

The time, $t$, is thus equal to $0$ for a newly divided cell and to $\tau$ at first passage, i.e., at division. The goal is to then compute the first passage time distribution, $F(\tau)$, where $\tau$ is the interdivision time, i.e., the time taken for the thresholded variable (cell size or function thereof) to reach the threshold value, $\th$. In this section, for additional clarity, we explicitly write out the parameters on which the FPT depends in its argument, separated from the variable, $\t$, by semi-colons. Thus, division at a threshold $\th$ has the FPT distribution: $F(\t;\th)$.

\subsubsection{Relation between cell sizes at division and interdivision times.}
\begin{figure}[ht]
\begin{center}
\resizebox{8.5cm}{!}{\includegraphics{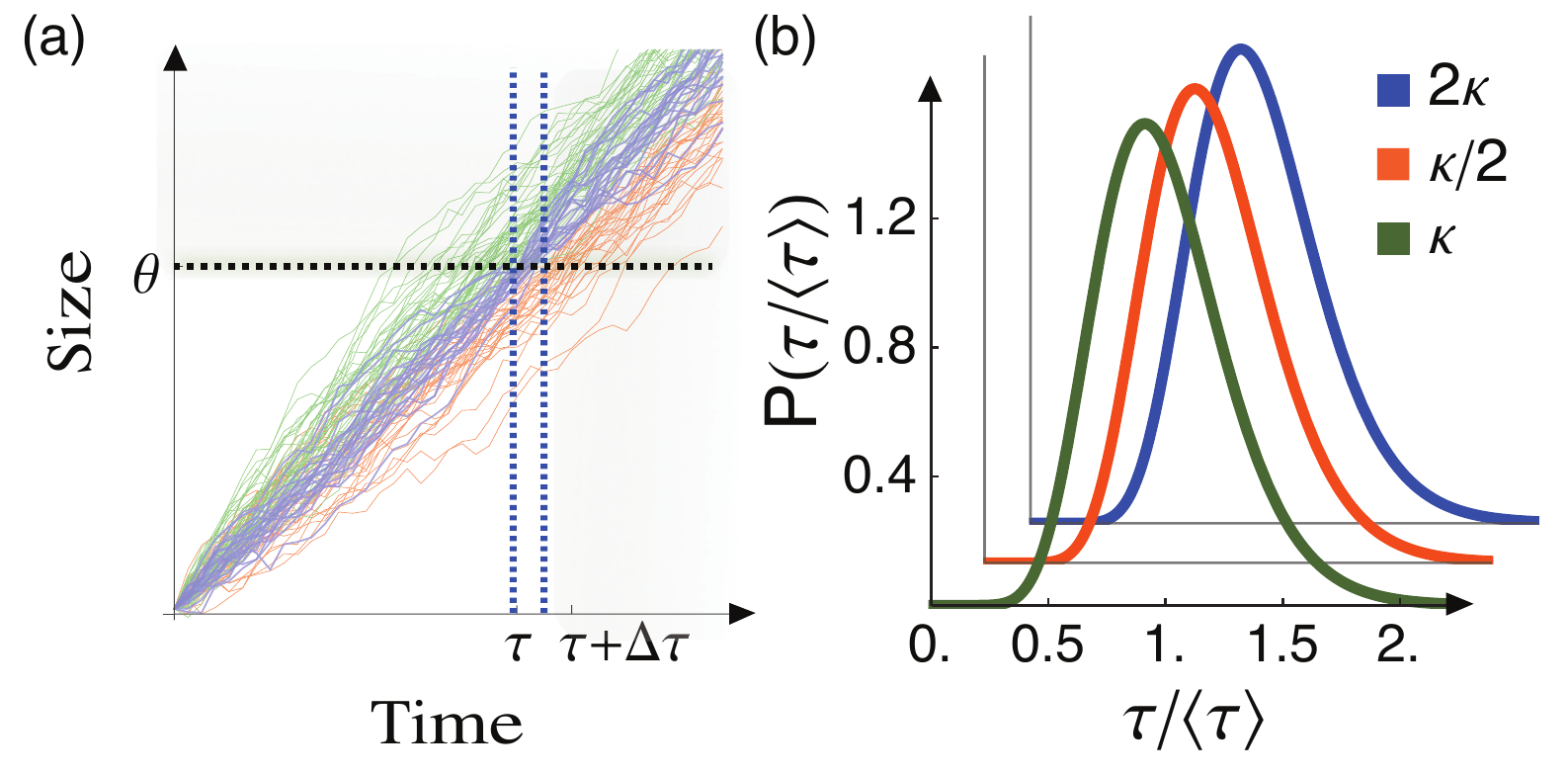}}
\caption{{\bf Cell division as a First Passage Time problem. (a)} Schematic of stochastic cell size increase from a common initial condition. Between times $\t$ and $\t + \D \t$, the  blue growth tracks cross the threshold size, $\th$. The green trajectories cross the threshold before this interval, and the red trajectories cross the threshold after this interval. Using probability conservation, the cumulative probability that the size is greater than $\th$ (above the black dotted horizontal line) must be equal to the complement of the cumulative probability that the First Passage Time is less than or equal to $\t$ (left of blue dotted vertical line at $\t$). {\bf (b)} Scaling of the First Passage Time distribution. The shape of the mean-rescaled division time distribution is timescale invariant, i.e., independent of $\k$, when there is a single timescale, $1/\k \propto \la\t\ra$, in the FPT dynamics.}
\label{fig-dsc}
\end{center}
\end{figure}

A convenient simplification of the FPT problem is obtained using a generic feature of cell growth: cell sizes always increase {\em monotonically} with time (for living cells), even though the increase is stochastic~\cite{-iyer-biswas-kv, -iyer-biswas-on}. The possibility of multiple crossings of the threshold is then automatically ruled out (since the threshold is passaged exactly once) and so we do not need to consider the auxiliary problem with absorbing boundary conditions (see Section \ref{sec-FPT-framework}). This leads to the following mathematical identity. Quite generally, when a stochastic variable $s$ increases monotonically with time, its time dependent distribution, ${P}(s,t)$, can be related to the distribution of First Passage Times, $F(\t; s= \th)$, through a simple geometric argument, which illustrates the derivations of Section \ref{sec-FPT-framework}. From Fig.~\ref{fig-dsc}, using probability conservation, it follows that the cumulative of the size distribution at the threshold value must be equal to the cumulative of the First Passage Time distribution. Thus the FPT distribution can computed using the relation:
\begin{align}
\label{eq-fpt}
F(\t; \th) =  \pd_{\t}\le[ \int_{\th}^{\infty} \; d s \,{P}(s, \t)\ri].
\end{align}
When a discrete growth model is used for $s$,  the integral should be replaced by an appropriate sum. In this section we denote cell size by $s$, irrespective of whether the stochastic growth model used is discrete or continuous.

\subsubsection{Scale invariance of the FPT distribution.}\label{sec-si}
Division time distributions from different growth conditions, for the same organism, have been observed to undergo scaling collapses, when rescaled by their condition-specific mean values~\cite{-iyer-biswas-kv, 2013-giometto-qf, 2014-kennard-xh}. This observation encodes a deeper truth about FPT distributions:  whenever a single timescale dominates the stochastic growth and division dynamics, irrespective of the functional from of the growth law, or the thresholding scheme, the mean-rescaled FPT distribution from different growth conditions is scale invariant.

To see how this result arises formally, we denote the assumed single timescale in the problem by  $\k^{-1}$ (in practice, this timescale can be tuned by external parameters). For the growth variable, $s$, whose time dependent distribution is ${P}(s, t)$, and $\th$ the threshold at which division (first passage) occurs, the FPT distribution is given by:
\begin{align}
F(\t; \th) &\equiv \pd_{\t} \le[ \int_{\th}^{{\infty}} ds\, {P}(s,\t; \k) \ri]  \nn \\
&= \k\, \pd_{(\k\, \t)} \le[  \int_{\th}^{{\infty}}  ds\, {P}(s, \k\, \t; \k =1) \ri].
\end{align}
Thus, if we now change variables to $\tilde{\t} = \k \t$ and look at its probability distribution, $\tilde{F}(\tilde{\t}; \th)$, then we have,
\begin{align}
\tilde{F}(\tilde{\t}; \th) = \frac{1}{\k} F(\t; \th)  =  \pd_{\tilde{\t}} \le[  \int_{\th}^{{\infty}}  d s\, {P}(s, \tilde{\t}; \k =1) \ri],
\end{align}
which is manifestly $\k$ independent.  Therefore, $\tilde{F}(\tilde{\t}; \th) $ is timescale-invariant and is the functional form of the scaling invariant mean-rescaled FPT distribution.

It is also straightforward to show that the mean FPT, $\la {\t} \ra$, is  proportional to $1/\k$.  Thus the mean-rescaled division time distribution from different growth conditions will be found to undergo a scaling collapse, provided the underlying stochastic growth model has only one timescale (as is true for exponential or linear growth), and the thresholding does not itself introduce new timescales into the FPT process (as assumed above).

Conversely, the observation of a scaling collapse of mean-rescaled FPT distributions confirms that a single timescale governs the underlying stochastic dynamics. While the formal result appears to be intuitively obvious, the implications of observing this in a real biological system are significant: the growth law, regardless of its functional form, must depend on just one timescale; the thresholding scheme does not itself introduce a new timescale into the division dynamics; the division and growth timescales must therefore be proportional to each other and the mean division time, as external parameters are changed.

The specific stochastic growth models that we consider in the following sections all have one timescale governing growth, and the thresholding schemes we have enumerated above are timescale independent. Thus, the division time  distributions for each case are scale-invariant, when mean-rescaled. (This can be checked directly from the analytical forms derived below.)
\subsubsection{Master Equation approach.}
In this section we find the FPT distribution using the Master Equation approach, by using discrete stochastic growth models for $s$ and the Master Equation framework (see Section~\ref{sec-forward-equations}). In the continuum limit of $s$, the results for the FPT distributions are essentially unchanged. As previously noted, for the purpose of the present discussion we assume that cell size growth is either linear of exponential.  (Also see ``phase oscillator model'' below for a different interpretation of the linear growth model.)

{\em Linear growth:} In a  simple stochastic (discrete) model for linear growth, represented by
\begin{align}
s &\eqwithrate{k} s + 1,
\end{align}
the time evolution of $s$ is governed by the Master Equation (see Eq.~\eqref{eq-masterequation0}):
\begin{align}
\pd_{t} P(s, t) = k \le [P(s-1, t) - P(s, t)\ri].
\end{align}
(See, for instance, ~\cite{gardiner-book-2003}.) We have made the standard assumption of exponentially distributed waiting times. Using standard techniques~\cite{gardiner-book-2003}, it is straightforward to show then that the ensemble mean, $\la s(t) \ra \equiv \sum_{s =0}^{\infty} s P(s, t)$, grows linearly with time as $k \,t$.

We now consider different scenarios for thresholding the size variable in this model. The initial condition is that the cell size distribution at $t=0$ is $R(s_{0})$, where $s = s_0$ at $t = 0$. We first consider the simplest initial condition: all cells start out with the same $s=0$ at $t=0$, i.e.,  $R(s_{0}) = \d_{s_{0}, 0}$. It can be shown that for this initial condition the time-dependent distribution of sizes at any given time  is then the Poisson distribution~\cite{gardiner-book-2003, 2009-iyer-biswas-ht} whose single (time dependent) parameter equal to both the mean and the variance of the Poisson distribution, is $k t$:
\begin{align}
{P}(s, t) = \frac{e^{- k t} \, {(k t)}^{s}}{s!}.
\end{align}
Now, for the absolute size threshold, $s= \th$, the FPT distribution is found by using equation
\begin{align}
F(\t; \th) =  \pd_{\t}\le[ \sum_{s=\th}^{\infty} {P}(s, \t)\ri].
\end{align}
Upon evaluation, we find that the FPT obey a Gamma distribution whose shape parameter is given by the magnitude of the threshold, $\th$:
\begin{align}
F(\t; \th; k)  =  \frac{k \,e^{- k \t}\, (k \t)^{-1+\th}}{\G[\th]},
\end{align}
where $\G[x]$ is the Gamma function~\cite{1980-gradshteyn-qf}, and we have explicitly written out the parametric dependences of $F$, as previously mentioned. Note that the FPT distribution $F(\t; \th; k)=kP(\theta -1,\t)$, which is the probability flux from state $\theta -1$ to state $\theta$, in accord with Section \ref{sec-FPT-framework}.

When {\em all} cells  are assumed to start with the same initial size, $s_{0}>0$, the size and division time distribution for absolute thresholding are, respectively, a shifted Poisson and a shifted Gamma distribution, as one might intuit:
\begin{align}
{P}(s, t; s_{0}) &=   \frac{e^{- k t} \, {(k t)}^{(s-s_{0})}}{(s-s_{0})!} \, \Th(s-s_{0}),
\end{align}
where $\Th(s-s_{0})$ is the Heaviside Theta function. The corresponding FPT distribution is  again found using~\eqref{eq-fpt}:
\begin{align}
F(\t; \th -s_{0}; k)  &=   \frac{k \, e^{- k \t}\, (k \t)^{-1+\th-s_{0}}}{\G[\th-s_{0}]}, \mbox{ for } \th > s_{0} +1 \mbox{ and }\nn \\
&= k \, e^{-k \t}, \mbox{ for } \th = s_{0} +1,
\end{align}
which is again identical to the probability flux into state $\theta$. A limiting case of the above solution, $\th = s_{0} +1$, is consistent with the assumption that the waiting time distribution is exponential. In this model the initial value, $s_0$, does not affect the propensity for stochastic growth. Thus, when there is an initial {distribution} of sizes, denoted by $R(s_{0})$, the resulting  FPT for absolute size thresholding is given by the convolution of the above Gamma distribution with $R(s_{0})$.

Next consider differential size thresholding, i.e., a cell with an initial  size $s_{0}$ divides when its size reaches $s = s_{0} + \Delta \th$, where the additive threshold, $\Delta \th$ is a given positive number. Using the result above for a given initial condition, $s_{0}$, we find now that the  division time distribution, for additive thresholding:
\begin{align}
F(\t; \Delta \th)  &=  \sum_{s_{0} = 0}^{\infty} R(s_{0})\times  \frac{k \, e^{-k \t}\, (k \t)^{-1+ \Delta \th}}{\G[\Delta \th]} =  \frac{k \, e^{-k \t}\, (k \t)^{-1+ \Delta \th}}{\G[\Delta \th]}, \mbox{ for } \Delta \th > 1, \nn \\
&= \sum_{s_{0} = 0}^{\infty} R(s_{0})\times k \, e^{- k \,\t} = k \, e^{-k \, \t} , \mbox{ for } \Delta \th = 1.
\end{align}
Not surprisingly, $R(s_{0})$ drops out of the expression, and the FPT distribution for this case is thus independent of the initial size distribution.

In all the above expressions for FPTs, we could set $k=1$, i.e., effectively measure all the times (including the division time, $\t$), in units of $1/k$. Evidently, when this substitution is made, the FPT distribution becomes timescale invariant. Moreover, in all cases, $\la \t \ra \propto 1/k$. Thus a scale-invariant result is obtained when the FPT distributions are rescaled by $\la \t \ra$, in agreement with the general scaling result derived previously in Section~\ref{sec-si}.

{\em Exponential growth:} A simple model in which the ensemble average of $s$ grows exponentially with time, represented by the growth process,
\begin{align}
s &\eqwithrate{k\,s} s + 1, \nn \\
\end{align}
undergoes time evolution governed by the Master Equation  (see Eq.~\eqref{eq-masterequation0}),
\begin{align}\label{eq-meshc}
\pd_{t} P(s, t) = k \le [(s-1) P(s-1, t) - s P(s, t)\ri].
\end{align}
As in the previous section, we first consider the initial condition where all cells have the same initial size, $s_{0}$, and then generalize to the case where they may have a distribution, $R(s_{0})$. We find that the cell size distribution is a Negative-binomial distribution~\cite{-iyer-biswas-on} (which is the discrete analogue of a Gamma distribution)~\cite{Karlis2005}:
\begin{align}
\;\;\;\;\;\;\;\;\,\,\,\;\;\;\;\;\,\,\;\;\;\;\;\;P(s, \t; s_{0}; k) & = {s-1 \choose s_{0}-1}(1- e^{-k \,\t})^{(s-s_{0})}\, e^{-k \,\t s_{0}}\Th(s-s_{0}).
\end{align}
The FPT for absolute size threshold $\th$ is then found using~\eqref{eq-fpt} to be the Beta-exponential distribution~\cite{2006-nadarajah-ly},
\begin{align}
F(\t; s_{0}, \th; k) &= \frac{k \, e^{-s_{0}k \,\t}(1- e^{-k\,\t})^{(-1 +\th -s_{0})}}{\b[s_{0}, (\th-s_{0})]},
\end{align}
where $\b[x,y]$ is the Beta-function~\cite{1980-gradshteyn-qf}. As before, it can be shown using the properties of the Beta-function that $F(\t; s_{0}, \th; k)=kP(\theta-1, \t; s_{0}; k)$, the probability flux into state $\theta$, in accord with Section \ref{sec-FPT-framework}. For an alternative approach leading to the Beta-exponential solution, see~\cite{1988-szabo-zr}. Note that the FPT distribution, which is a Beta-exponential in $\t$, is actually a Beta distribution if one transforms the variable $\t$ to $\n \equiv e^{-k \,\t}$. The parameters of this Beta distribution are restricted such that the FPT distribution is always unimodal in this problem.

We now formulate the {ratio} thresholding problem for exponential growth. Starting with an initial size, $s_{0}$, drawn from an initial size distribution, $R(s_{0})$, each cell is assumed to divide when its size reaches a ratio threshold, $s(\t)/s_{0} \equiv r$. Upon solving the Master Equation~\eqref{eq-meshc} we get:
\begin{align}
{P}(s, t; k) &= \sum_{s_{0} = 0}^{\infty} \,R(s_{0})\times\, {s-1 \choose s_{0}-1}(1- e^{-k \,t})^{(s-s_{0})}\, e^{-k \,ts_{0}}\Th(s-s_{0}).
\end{align}
Therefore, using Eq.~\eqref{eq-fpt} again, the FPT  using a ratio threshold, $r$, is
\begin{align}
F(\t; r; k) &= \sum_{s_{0} = 0}^{\infty} \,R(s_{0})\times\, \frac{k \, e^{-s_{0}k \,\t}(1- e^{-k\,\t})^{(-1 +r s_{0} -s_{0})}}{\b[s_{0}, (r s_{0}-s_{0})]},
\end{align}
whose shape  {depends on the specific choice of initial size distribution, $R(s_{0})$}.

We note that it may be possible to invoke overarching biophysical principles constraining growth and division in population balance to self-consistently determine the initial size distribution, $R(s_{0})$. However, this discussion is outside the scope of this review.

To recapitulate, we have shown in this subsection how given a (mean) growth law and a thresholding scheme for division, the FPT problem for cell division can be formulated and the cell division time distribution can be analytically derived. Such a model may be used to make other predictions about the biological system, including placing constraints on the possible  topologies of the networks governing the stochastic growth~\cite{-iyer-biswas-on}. All division time distributions derived above, for different growth laws and thresholding schemes, are positively skewed (have a long right tail), and are unimodal. However, it is worth mentioning that even with the kind of high quality data available in recent single-cell experiments~\cite{-iyer-biswas-kv, 2010-wang-cr}, using the shape of the observed division time distribution to infer the underlying growth law (for example, to distinguish between linear and exponential growth) is extremely challenging and not practically feasible.

\subsubsection{Cell cycle as a phase oscillator.}
In some scenarios the cell cycle (i.e., the intervening period between successive divisions) is modeled as a phase oscillator, with the cell cycle phase, $\phi$ being set equal to  $0$  for a newly divided cell and $\phi = 2 \pi$ for a cell about to divide~\cite{1991-cooper-vc, 2010-lin-fk, 1987-alt-ew, 1989-tyson-ez}. If $N$ sequential steps need to be completed, as the cell cycle phase increases from $0$ to $ 2 \pi$ in steps of $2 \pi/N$, and if the waiting time distribution for each step is exponentially distributed as $k \exp(-k t)$, then the FPT problem for cell division in this model is essentially identical to the one for the stochastic discrete linear growth model solved previously. Thus the FPT distribution is the gamma distribution:
\begin{align}
F(\t; N; k)  &=   \frac{k \, e^{- k \t}\, (k \t)^{N - 1}}{\G[N]}, \mbox{ for } N >1 \mbox{ and }\nn \\
&= k \, e^{-k \t}, \mbox{ for } N = 1.
\end{align}
In principle, the observed cell division time distributions can therefore be fitted to a gamma distribution, and used to estimate the number of ``elementary'' steps in the cell cycle, $N$. However, as we have shown above, the model makes the simplistic assumption, almost certainly violated by all realistic systems, that all steps have the identical and exponentially distributed waiting-time statistics. Thus, the practical utility of such an estimate of ``N'' is limited.

\subsubsection{Phenomenological approach using Langevin or Fokker-Planck frameworks.}
In the examples that we have considered thus far, we have used a microscopic model to motivate the growth law, and then used it to find the FPT distribution. A complementary approach is the phenomenological one in which one proposes Langevin dynamics consistent with the observed mean growth law, and assumes an ansatz for the noise term for cell size fluctuations. By then going to the corresponding Fokker Planck description, one can use standard techniques~\cite{redner-book} (see Section \ref{sec-FPT-framework}) to compute the FPT distribution.

We elucidate this alternative approach with the specific case of the exponential growth law, with ratio size thresholding. This methodology can be readily adapted to other growth laws and thresholding schemes.

The cell size increases, on average, as $\la s(t) \ra  =  \la s(0) \ra \exp(k t)$; the cell divides at a time,
 $\t$, such that $s(\t)/ s(0)$ is a constant. Motivated by the exponential growth law and ratio thresholding scheme assumed, we define a new stochastic dynamical variable for each cell, $x(t)$, as:
\begin{align}\label{eq-alpha}
x(t) \equiv  \log\le[{s(t)}/{s(0)}\ri],
\end{align}
where $s(0)$ is the initial size of the cell under consideration. The threshold for division is then given by
\begin{align}\label{eq-thresh}
x_{o} \equiv x(\t) =  \log\le[\frac{s(\t)}{s (0)}\ri].
\end{align}
We then write a stochastic growth model as a Langevin equation (see Eq.~\eqref{eq-SDE-Ito}) for the ``Brownian motion'' of $x(t)$:
\begin{align}\label{eq-EOM}
\frac{d \,x(t)}{d t} = \k + \sqrt{B}\,\, \chi(t),
\end{align}
where $\k$ is the mean (ensemble averaged) exponential growth rate, the ``drift'' term; the second term  is the `noise' term: $\chi(t)$ is  standard Gaussian white-noise, and the ``diffusion'' term $\sqrt{B}$ measures the strength of the noise in $x$. To evaluate the FPT distribution, we first recast this Langevin equation to its equivalent Fokker Planck equation, Eq. \ref{eq-fokkerplanck-SDE} and Eq.~\eqref{eq-fokkerplanck-B(x)}, using standard techniques. It is:
\begin{align}\label{eq-DE}
\pd_{t} {P}(x, t) + \k\,\, \pd_{x} {P}(x, t) = \frac{B}{2}\,\, \pd_{x}^{2}{P}(x,t).
\end{align}
For computational ease, without loss of generality, we shall assume that each cell (the ``random walker'') starts out at $x = x_{0}$ at $t=0$ and divides (gets absorbed) when it first crosses the origin, $x =0$. Thus our initial condition is that ${P}(x, t=0) = \d(x - x_{0})$ and we shall impose absorbing boundary condition at $x = 0$. The fraction of random walkers disappearing at $x=0$ between times $\t$ and $\t + d\t$ is then related to the current of walkers entering  $x=0$ in that time interval according to the probability conservation equation, Eq.~\eqref{eq-22},
\begin{align}\label{eq-FPT}
F(\t)\, d\t &= \int_{0}^{\infty} \le[{P}(x, \t) - {P}(x, \t + d\t)\ri] \, dx \nn \\
&= \le( -\frac{\pd}{\pd \t} \int_{0}^{\infty} {P}(x, \t) \, dx \ri)\, d\t,
\end{align}
where $F(\t)$ is the First Passage Time distribution sought. We note that ${P}(x, t)$ is ``normalized'' at each time such that
\begin{align}
\int_{0}^{\infty} d x \, {P}(x, t) = 1 - \int_{0}^{t} d\t \, F(\t),
\end{align}
while the FPT, $F(\t)$, is itself correctly normalized to $1$. This is because ${P}(x, t) $ quantifies the density of the surviving random walkers (see Sections \ref{sec-FPT-framework} and \ref{sec-kramers}).

The solution to the Fokker Planck equation \eqref{eq-DE}, with the specified boundary conditions, is most elegantly computed  using the method of images, which is routinely used for electrostatics problems with symmetry~\cite{1999-jackson-zh, redner-book}. We note that, contrary to naive expectation,  the random walker and its image must move in the {\em same} direction to satisfy the Fokker Planck equation. The solution, which can be verified straightforwardly by substitution, is
\begin{align}\label{eq-DE-soln}
{P}(x, t) = \frac{1}{\sqrt{2\pi B\,t}}e^{-(x -x_{0} + \k \, t)^{2}/2 B\,t} ( 1- e^{-2\,x \, x_{0}/B\,t}).
\end{align}
Using this expression for ${P}(x, t)$ and Eq.\eqref{eq-FPT}, we find the First Passage Time distribution,
\begin{align}
F(\t) = \frac{x_{0}}{\sqrt{2\pi B\,\t^{3}}} e^{-(x_{o} - \k\,\t)^{2}/2B\,\t},
\end{align}
which is the so called  {Inverse Gaussian Distribution}, with parameters $\k$ and $B$. It was first derived as the FPT distribution for Brownian motion by Schrodinger~\cite{1915-schrodinger-ly}.

Notably, the shape of the Inverse Gaussian Distribution {``scales''} with the mean, as expected (see preceding section on scaling of FPT distributions). As previously noted, cell sizes are observed to increase monotonically, even when fluctuations are considered. This implies that for the phenomenological descriptions such as presented above (which assume Gaussian white noise), the drift term must be overwhelmingly larger than the diffusive term, resulting in (approximately) monotonic  growth. In this drift dominated regime, the P\'eclet number for this FPT problem~\cite{redner-book} is thus a very large dimensionless number.
For this model, it is equal to $1/\eta^{2}$, where $\eta$ is the coefficient of variation of  $F(\t)$. In this regime $\k \,\a_{0} \gg \sqrt{B \,\a_{0}}$ and so  the ``distance'' travelled by the peak of the distribution of $x$ in a given time interval is much greater than the corresponding widening of the distribution. In the drift dominated limit the problem is approximately equivalent to the motion of a Gaussian probability packet, ${Q}(x,t)$, which is the solution to the same problem with the same initial condition, but with no absorbing boundary condition at $x = 0$. The error in this approximate solution comes from the fact that we do not take into account  that the same particle might have crossed the boundary at $x = 0$ more than once.

We note that  parameters of this phenomenological description can be inferred  from experimental observations.  The drift, $\k$, is directly given by the ensemble-averaged growth curves. From  time series growth data, the mean squared displacement of $x$ with time can be computed to confirm that the behavior is diffusive, and to read off the diffusion strength, $B$. The scale invariance of the FPT, when external parameters are tuned, provides an additional check on whether these two parameters are then found to be related to each other as predicted.

\section{Concluding remarks.}
Recent years have seen a renewed interest in stochastic processes in biology, including First Passage processes, resulting in a rapidly increasing wealth of literature. The aims of this review were two-fold. First, we consolidated the theoretical foundations and techniques of FP processes within a unified framework. Second, we provided an introduction to the practical use of FP methods in biophysical applications using several pertinent examples.

Out of necessity, the applications discussed here do not constitute an exhaustive list of potential uses of the FP theory, even just in the cellular context. We apologize to the authors whose work could not be cited. However, the methods and techniques discussed here have been successfully applied to molecular motors,  translation, transcription, protein and enzyme dynamics as well as signaling. Beyond the context of cell biology, vast literature on FP applications to neurobiological systems and population genetics can be found. We point the interested reader to some relevant literature on these topics~\cite{klumpp-hwa-translation-2008,2014-metzler-dq,motor-review-chowdhury-2013,Kolomeisky2011,kolomeisky-fisher-2007-motors-review,Hanggi2009,whitford-protein-dynamics-review-2012biomolecular,kinetic-proofreading-coombs-goldstein-2005,Bressloff-Review-2014,chou-review-2014,meerson-review-2010,enzymes-times-xie-2005,bel-munsky-kinetic-proofreading-2010,nemenman-receptor-times-2013}.

Several important theoretical aspects also were not reviewed here, most notably those pertaining to systems with fluctuating barriers and boundaries \cite{doering-fluctuating-barrier-1992,chou-nowak-dynamic-2007,magnasco-fluctuating-2013}, anomalous diffusion \cite{klafter-anomalous-review,barkai-metzler-anomalous-review-2014,bel-barkai-2005weak} and Hamiltonian methods for large fluctuations \cite{meerson-review-2010,Bressloff-Review-2014}. Finally, in this review  we focused entirely on the analytical approaches. Simulation techniques that complement analytical approaches have played a crucial role in revealing the stochastic mechanisms of cells and molecules. We direct the reader to the following textbooks and reviews as starting points for further inquiry~
\cite{2007-gillespie-cr, 2014-metzler-dq,gillespie-book-1991}.

\section*{Acknowledgements.} We thank Ariel Amir, Matthew Badali, Golan Bel, Sasha Berezhkovskii, Tom Chou, Gavin Crooks, Sean Crosson, Aaron Dinner, Stefano Di Talia, Aretha Fiebeg, Jon Henry, Leo Kadanoff, Marcelo Magnasco, John Pearson, Sidney Redner, Norbert Scherer, Stas Shvartsman, Ze'ev Schuss, David Sivak, Brian vanKoten and Charlie Wright for useful discussions and Matthew Badali, Rudro Biswas, Joshua Milstein and Meni Wanunu for careful reading of the manuscript. A.Z. acknowledges support from Canada National Science and Engineering Research Council. S.I.-B. was supported by the National Science Foundation (NSF PHY-1305542) and the W. M. Keck Foundation.

\end{document}